\documentclass[3p, preprint]{elsarticle}

\journal{Physical Review D}

\usepackage[english]{babel}
\usepackage{units}
\usepackage{graphicx}
\usepackage{amsmath}
\usepackage{amssymb}
\usepackage{caption}

\usepackage{xcolor}
\usepackage{xspace}

\usepackage{pgf}
\usepackage{caption}
\usepackage{subcaption}

\definecolor{dark-gray}{gray}{0.3}

\newcommand{\Xmax}{$X_{\rm max}$\xspace}
\newcommand{\Xmaxmath}{X_{\rm max}}

\begin{document}

\title{Depth of shower maximum and mass composition of cosmic rays from 50~PeV to 2~EeV measured with the LOFAR radio telescope}

\author[br,ru]{A.~Corstanje\corref{cor}}
\cortext[cor]{Corresponding author}
\ead{A.Corstanje@astro.ru.nl}
\author[br,ru]{S.~Buitink}
\author[ru,as,ni]{H.~Falcke}
\author[kap]{B.~M.~Hare}
\author[ru,ni,br]{J.~R.~H\"orandel}
\author[ikp,br]{T.~Huege}
\author[br]{G.~K.~Krampah}
\author[br]{P.~Mitra}
\author[br]{K.~Mulrey}
\author[desy,ecap]{A.~Nelles}
\author[br]{H.~Pandya}
\author[br]{J.~P.~Rachen}
\author[iuibr]{O.~Scholten}
\author[as]{S.~ter Veen}
\author[abu]{S.~Thoudam}
\author[viet]{G.~Trinh}
\author[mpifr]{T.~Winchen}

\address[br]{Astrophysical Institute, Vrije Universiteit Brussel, Pleinlaan 2, 1050 Brussels, Belgium}
\address[ru]{Department of Astrophysics/IMAPP, Radboud University Nijmegen, P.O. Box 9010, 6500 GL Nijmegen, The Netherlands}
\address[as]{Netherlands Institute for Radio Astronomy (ASTRON), Postbus 2, 7990 AA Dwingeloo, The Netherlands}
\address[ni]{Nikhef, Science Park Amsterdam, 1098 XG Amsterdam, The Netherlands}
\address[kap]{University of Groningen, Kapteyn Astronomical Institute, Groningen, 9747 AD, Netherlands}
\address[ikp]{Institut f\"{u}r Astroteilchenphysik, Karlsruhe Institute of Technology (KIT), P.O. Box 3640, 76021, Karlsruhe, Germany}
\address[desy]{DESY, Platanenallee 6, 15738 Zeuthen, Germany}
\address[iuibr]{Interuniversity Institute for High-Energy, Vrije Universiteit Brussel, Pleinlaan 2, 1050 Brussels, Belgium}
\address[ecap]{ECAP, Friedrich-Alexander-University Erlangen-N\"{u}rnberg, 91058 Erlangen, Germany}
\address[abu]{Department of Physics, Khalifa University, P.O.~Box~127788, Abu Dhabi, United Arab Emirates}
\address[viet]{Department of Physics, School of Education, Can Tho University Campus II, 3/2 Street, Ninh Kieu District, Can Tho City, Vietnam}
\address[mpifr]{Max-Planck-Institut f\"{u}r Radioastronomie, Auf dem H\"ugel 69, 53121 Bonn, Germany}

\begin{abstract}
We present an updated cosmic-ray mass composition analysis in the energy range $10^{16.8}$ to $\unit[10^{18.3}]{eV}$ from 334 air showers measured with the LOFAR radio telescope, and selected for minimal bias. In this energy range, the origin of cosmic rays is expected to shift from galactic to extragalactic sources.
The analysis is based on an improved method to infer the depth of maximum \Xmax of extensive air showers from radio measurements and air shower simulations.

We show results of the average and standard deviation of \Xmax versus primary energy, and analyze the \Xmax-dataset at distribution level to estimate the cosmic ray mass composition. Our approach uses an unbinned maximum likelihood analysis, making use of existing parametrizations of \Xmax-distributions per element. The analysis has been repeated for three main models of hadronic interactions.

Results are consistent with a significant light-mass fraction, at best fit $23$ to $\unit[39]{\%}$ protons plus helium, depending on the choice of hadronic interaction model. The fraction of intermediate-mass nuclei dominates. This confirms earlier results from LOFAR, with systematic uncertainties on \Xmax now lowered to 7 to $\unit[9]{g/cm^2}$. 

We find agreement in mass composition compared to results from Pierre Auger Observatory, within statistical and systematic uncertainties. 
However, in line with earlier LOFAR results, we find a slightly lower average \Xmax. The values are in tension with those found at Pierre Auger Observatory, but agree with results from other cosmic ray observatories based in the Northern hemisphere. 
\end{abstract}

\maketitle

\section{Introduction}
Cosmic rays arrive at the Earth's atmosphere in an energy range from below $10^9$ to above $\unit[10^{20}]{eV}$.
Upon interacting in the atmosphere, they produce a cascade of secondary particles called {\it extensive air shower}, which is measurable in ground-based detector arrays for energies above about $\unit[10^{14}]{eV}$. 
At the high end of the energy spectrum, these particles have the highest energy of the known particles in the Universe. 
Therefore, the questions about their origin and their mass composition have raised considerable interest, and cosmic-ray air showers are measured in observatories around the world. The largest is the Pierre Auger Observatory in Argentina \cite{Abraham:2010,Abraham:2004}, spanning an area of $\unit[3000]{km^2}$.

In this analysis, we study cosmic rays with a primary energy between $10^{16.8}$ and $\unit[10^{18.3}]{eV}$, the energy range where a transition is expected from particles originating from within the Galaxy, to an extragalactic origin.
Heavy nuclei from the Galaxy are expected to reach higher energies than protons, as they are more easily magnetically contained due to their higher charge (i.e.,~the Hillas criterion \cite{Hillas:1984}). 
Therefore, composition measurements in this energy region are interesting for comparison with models of cosmic-ray sources and propagation. 
For instance, in \cite{ThoudamAandA:2016} it is argued that a secondary Galactic component may (still) dominate around $\unit[10^{17}]{eV}$. In one scenario where supernovas of Wolf-Rayet stars are the main sources, one expects a rather low proton fraction together with a higher helium and C/N/O fraction, before proton-dominated extragalactic cosmic rays take over around $\unit[10^{18}]{eV}$.

Along the track of an air shower, the number of secondary particles reaches a maximum, at a depth expressed in $\unit{g/cm^2}$ of traversed matter, referred to as \Xmax. This maximum is reached for almost all showers in our energy range, typically at altitudes of 2 to $\unit[7]{km}$.
At a given primary energy, \Xmax depends on the mass of the primary particle. It is different for protons compared to heavy nuclei, both on average and in distribution.
The shift in average \Xmax with respect to protons is approximately proportional to $\ln A$, for particles with atomic mass number $A$, where protons have the deepest shower maximum on average.
Thus, measuring \Xmax for a collection of air showers gives information about their composition, and is the basis for the present analysis.

There are three main techniques for measuring \Xmax:~(i) measuring fluorescence light along the trail of the air shower, (ii) measuring Cherenkov light, and (iii) measuring the radio signal using antennas on the ground \cite{Kampert:2012, Huege_review:2016}. Measuring secondary particles on the ground, especially the electron/muon ratio, yields composition information without (explicitly) measuring \Xmax.
The radio detection technique has shown substantial development in recent years, leading to a method to determine \Xmax with a resolution about $\unit[20]{g/cm^2}$ \cite{Buitink:2014}. The method has been demonstrated using the LOFAR radio telescope, showing that the cosmic rays around $\unit[10^{17}]{eV}$ have a considerable light-mass component \cite{Buitink:2016}. 
Here, we present a method which has been improved on several points, thus lowering the systematic uncertainties, and an extended dataset.

The method relies on air shower simulations tracking individual particles, and summing up their contributions to the radio signal measured on the ground. For this, the CORSIKA \cite{Corsika:1998} simulation program has been used, with its plugin CoREAS \cite{CoREAS:2013} for computing the radio signal.
For an ensemble of simulated air showers, their lateral intensity distribution or `radio footprint' is fitted to the measurements, from which \Xmax and the energy of the measured shower are reconstructed.

The \Xmax-distributions for the different elements have substantial overlap.
Achieving low systematic uncertainties on \Xmax is therefore a crucial point for composition measurements, besides a good \mbox{\Xmax-resolution} per shower. This is done by a fiducial sample selection based on the CoREAS simulations per shower, and lowering known contributions to systematic uncertainties where possible. For example, in \cite{Corstanje:2017} it was shown that accurately representing local atmospheric conditions (refractive index) at the time of the air shower, removes a systematic error of 4 to $\unit[11]{g/cm^2}$. 

Other improvements to the analysis include a radio-only reconstruction of both \Xmax and energy, the latter using a new calibration based on Galactic emission \cite{Mulrey:2019} which halves the systematic energy uncertainty compared to the earlier particle-based treatment. Using a fast pre-computation of shower simulations with CONEX \cite{Bergmann:2007} streamlines the reconstruction, as showers can be pre-selected for their \Xmax.
The selection criteria to obtain a bias-free \Xmax sample have been improved, and a refined statistical analysis has been done.
All these increase the accuracy of the composition analysis, by lowering systematic and/or statistical uncertainties.

The Low Frequency Array (LOFAR) \cite{vanHaarlem:2013} is a radio telescope consisting of many separate antennas. 
The core region in the north of the Netherlands has a high density of antennas. The antennas are grouped in stations, each of which in the Netherlands contains 96 low-band antennas (LBA), working in the 10~to~$\unit[90]{MHz}$ range, and 48 high-band antennas (HBA) operating at $110-\unit[240]{MHz}$. The center of LOFAR is a circular area of $\unit[320]{m}$ diameter, with six of those stations. In a core region of about $\unit[6]{km^2}$, there are 18 more stations. 
LOFAR uses ring buffers to store up to 5 seconds of the raw measured signals at each antenna, which are used to measure the radio signals of air showers.
For air shower measurements, we use signals from the low-band antennas, filtered to 30 to $\unit[80]{MHz}$.

To trigger a buffer readout when an air shower arrives, a particle detector array called LORA (LOFAR Radboud Air shower Array) \cite{Thoudam:2014} is located inside the innermost ring of LOFAR. With 20 scintillator detectors monitored in real time, a trigger is sent to LOFAR when a threshold of 13 coincident detections is reached, a level which is optimal for our purposes.

The paper is organized as follows: in Sect.~\ref{sect:method}, we present the method of fitting air shower simulations to measured data to infer \Xmax. Furthermore, we discuss the selection criteria used to obtain a bias-free sample of showers. 
In Sect.~\ref{sect:composition}, the statistical analysis to infer particle composition from the \Xmax values is explained. 
The results are split into two sections, Sect.~\ref{sect:results_xmax} for the \Xmax distribution from our dataset, and Sect.~\ref{sect:results_composition} for the composition results. 
A summary is given in Sect.~\ref{sect:summary}.

\section{Method}\label{sect:method}
The discussion of the methods is split into five sections. After an introduction to the use of CORSIKA and CoREAS simulations, we give a brief review of the procedure to infer \Xmax for individual measured air showers. A more detailed explanation is found in \cite{Buitink:2014}; the details that have changed in this version are given below. The method to estimate the primary energy, and its uncertainties, are discussed in Sect.~\ref{sect:energy}. 
We show how including the local atmospheric conditions into the simulations leads to improved accuracy.
Finally, we explain our method to select showers in order to create an unbiased sample.

\subsection{CORSIKA/CoREAS simulations}
For the reconstruction of LOFAR-measured air showers, we use CORSIKA (version 7.7100) to simulate air showers, with its plugin CoREAS which calculates the radio emission from the particle showers.
The simulation uses a `microscopic' approach: it simulates individual particles and their contribution to radio emission, as they are produced along the evolution of the shower. Air showers are simulated from a Monte Carlo approach to particle interactions, and applying classical electrodynamics (Maxwell's equations) to obtain the radio signal at the antennas.
For the particle part, three main models of hadronic interactions have been considered: QGSJetII-04 \cite{Ostapchenko:2013}, EPOS-LHC \cite{EPOSLHC:2013}, and Sibyll-2.3d \cite{Sibyll:2020} \cite{Sibyll:2009}.
Their differences represent the intrinsic (systematic) uncertainty on hadronic interactions at the high energy levels of cosmic rays in our energy range and beyond.

The radio signals are calculated from first principles, i.e.,~without free parameters other than those from discretization approaches, which are set to values fine enough to reach convergence in results. This is important for accuracy in reconstructing \Xmax. Calculations are based on the `endpoint formalism' presented in \cite{James:2011}. In particular, there is no distinction between separate emission mechanisms such as geomagnetic and charge-excess contributions (see e.g.~\cite{Scholten:2008}), as these are naturally included.
The radio signals at ground level are a (coherent) superposition of contributions from particles along the shower track, propagated geometrically to the antennas. Therefore, one reconstructs essentially the (geometric) distance to \Xmax from the radio signals.

When fitting simulated air showers to LOFAR measurements, close agreement is found, for pulse energy \cite{Buitink:2016} as well as for detailed measurements such as circular polarization \cite{Scholten:2016}. 
Comparisons of results from Corsika/CoREAS with another simulation program based on the same principles, ZHAireS \cite{ZHAireS:2012}, show close agreement \cite{Huege_review:2016,Gottowik:2018,Mulrey:2020}, and remaining differences are ascribed to details in the simulation of particle interactions.
The given detailed and, where possible, parameter-free approach, together with agreement between different extensively developed simulation codes, gives a solid basis for accurate air shower reconstructions.  

\subsection{Using CoREAS simulations to estimate \Xmax of measured air showers}\label{sect:method_xmax}

Starting point is a set of air showers measured with LOFAR. When an event is triggered by the particle detector array LORA, its radio dataset is passed through our analysis pipeline \cite{Schellart:2013}. Its primary output parameter for this analysis is pulse `energy', defined as the square of the measured voltage in an antenna, integrated over a time window of $\unit[55]{ns}$ (11 samples) around the pulse maximum.
When reconstruction quality criteria are passed, a dataset consists of pulse energy, including its uncertainty, per antenna in at least 3 LOFAR stations, an accurate measurement of the incoming direction, and an initial estimate of \Xmax and primary energy from fitting a (parametrized) lateral distribution function \cite{Nelles:2015}. The initial estimates are used as a starting point for the simulations. Simulations are iterated if the initial parameters are found to be inaccurate. Therefore, the final estimates do not depend on them.

For each shower measured with LOFAR, we produce an ensemble of CoREAS showers, spanning the natural range of \Xmax for protons and for iron nuclei. 
We use the QGSJetII-04 hadronic interaction model \cite{Ostapchenko:2013} to produce the particle showers with CORSIKA.
As simulation energy we use an estimate from fitting an analytic description of the radio footprint \cite{Nelles:2015}, or an estimate from the particle detectors when the fit failed to converge. 

As a pre-computation stage we produce 600 showers with the fast simulation method CONEX, version 4.3700 \cite{Bergmann:2007}, of which 150 have an iron primary while the others start from a proton.
This is suitable to select those random number seeds to span the natural range of \Xmax roughly uniformly with about 15 showers. The same random number seeds are used in the full CORSIKA (version 7.7100) simulations. The number of CONEX showers is high enough to sample into the tails of the \mbox{\Xmax-distributions} at a level corresponding to the size of our final dataset ($N=334$). 
The aim is twofold, to have simulated showers covering the entire range of \Xmax, which is important for the selection criteria for a bias-free sample (see Sect.~\ref{sect:biasfree}), and to have a region around the best-fitting \Xmax with extra dense coverage, to improve precision.

\begin{figure}[h]
\begin{center}
\includegraphics[trim={3cm 0 3cm 0}, width=1.00\textwidth]{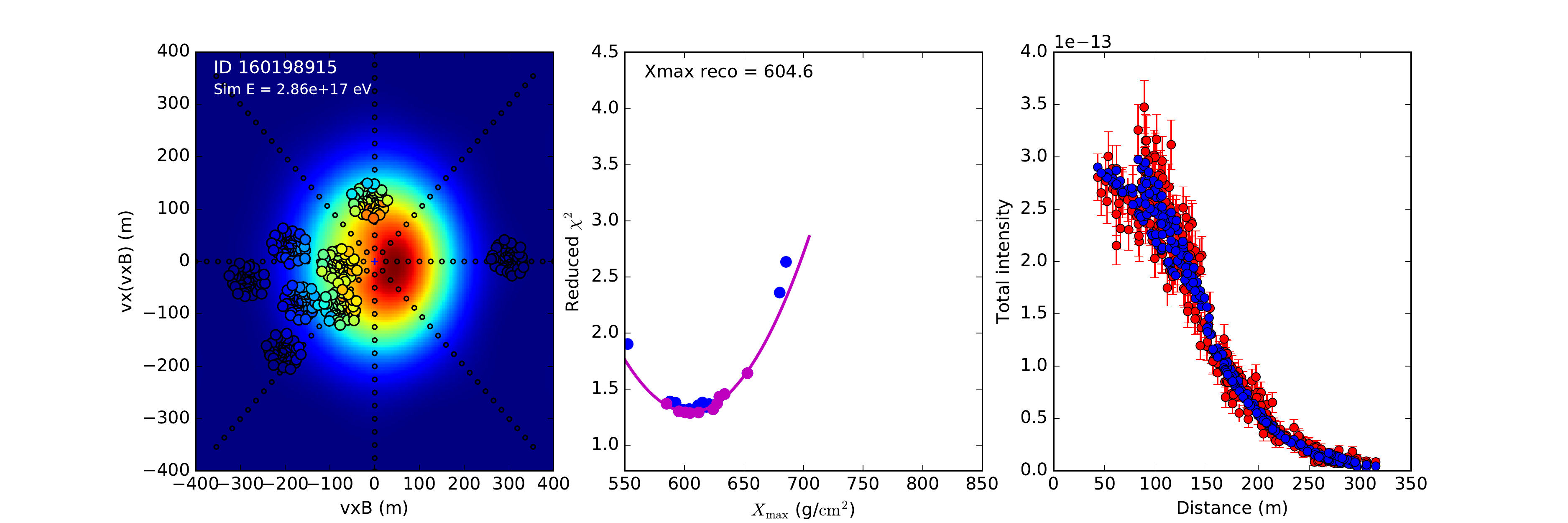}
\caption{
An example of fitted CoREAS showers to a LOFAR-measured shower. The left panel shows simulated signal energy along with the measurements, in the shower plane, for the best-fitting shower. The lateral intensity (and pulse energy) distribution is not rotationally symmetric. The middle panel shows the reduced $\chi^2$ as a function of \Xmax, with a parabolic fit through the lower envelope denoted by the magenta line. The right panel shows a 1-D lateral distribution function, where red points with uncertainties denote the measurements, and blue points denote the simulated intensities. }
\label{fig:event_fit_example}
\end{center}
\end{figure}

Therefore, ten additional showers are simulated in a region of $\pm \unit[20]{g/cm^2}$ around the first \Xmax estimate, aiming to have a high density of simulations close to the reconstructed \Xmax. The total number of simulated showers is around 30 per measured shower. If the reconstructed \Xmax deviates from the initial fit, extra showers are simulated to match the dense region with the reconstructed \Xmax. An example is shown in the middle panel of Fig.~\ref{fig:event_fit_example}; existing showers with $\Xmaxmath > 700$ fall outside the plotted vertical range.

The radio signal of each simulated shower is passed through our antenna model for the LOFAR LBA antennas \cite{Schellart:2013}, and through the bandpass filter used in the data analysis, to be able to compare with LOFAR data.
The signal energy for each simulated shower is then matched per antenna to the LOFAR measurements. 
In this fit, the core position and an overall scaling factor are free parameters.
This gives a chi-squared value for each shower:
\begin{equation}
\chi^2_{\rm radio} = \sum_{\rm antennas} \left(\frac{P_{\rm ant} - f_r^2\,P_{\rm sim} \left(x_{\rm ant} - x_0,\,y_{\rm ant} - y_0\right)}{\sigma_{\rm ant}}\right)^2,
\end{equation}
where $P_{\rm ant}$ and $\sigma_{\rm ant}$ denote the measured signal energy and its uncertainty, and $P_{\rm sim}$ is the simulated pulse energy. The overall scaling factor is $f_r^2$, and $(x_0, y_0)$ is the fitted shower core position.
In contrast to the method in \cite{Buitink:2014}, we perform the fit based on the radio signals only, making the radio reconstruction self-sufficient. In the previous analysis, the fit included both radio and particle detector data.
As a consequence, showers for which the reconstruction cannot be done accurately without the particle detector signals are now (automatically) discarded.

The result of the fitting procedure for one of our measured showers is shown in Fig.~\ref{fig:event_fit_example}. 
In the left panel, the best-fitting simulated shower is shown (background color) together with the measurements (colored circles).
The colored circles blend in well with the background color, indicating a good fit. 
This is confirmed by the middle plot, showing a reduced $\chi^2$ of 1.3 for the best fit, and a clear minimum as a function of \Xmax.
The right panel shows a one-dimensional representation of the simulated and measured intensities per antenna.

We employ a Monte Carlo procedure, using the simulated ensemble of showers to infer the uncertainties on \Xmax, the energy and the shower core position.
For each simulated shower in our ensemble we create three mock datasets as they would have been measured, i.e.,~adding the noise level found in each LOFAR antenna. They represent three different realisations of the random noise, at a fixed shower core position.
This simulated shower is then reconstructed through the above procedure, using the ensemble of all other simulations. Statistically comparing the reconstructions with the real \Xmax, core position, and radio scale factor, which are known in simulations, yields their uncertainties. The uncertainties thus found are calculated from the entire simulated ensemble, and are applicable to the measured shower as well as to each simulated shower; this is important in the bias-free sample selection procedure explained in Sect.~\ref{sect:biasfree}.
This procedure, relying on the reconstruction method described above, now also uses only the radio signals.

\subsection{Estimate of the primary particle energy}\label{sect:energy}
We estimate the energy of the primary particle by comparing the pulse energy of the measured radio signal with the predicted radio signal from CoREAS, which was produced at a given simulation energy obtained from the initial fit.
The intensity, and measured pulse energy, of the radio signal scale quadratically with the primary energy \cite{PAO:2016,Nelles_param:2015,Glaser:2016}.
Fitting CoREAS radio signal energy to LOFAR data produces an overall scale factor. The square root of this is taken as a correction to the simulation energy, giving an estimate of the primary particle energy. Moreover, following \cite{Gottowik:2018} we apply a correction factor of $\unit[11]{\%}$ to the simulated pulse intensities. This accounts for a finite step size (electron multiple scattering length) in tracking the particle cascade in the simulation.

This procedure of matching simulated and measured pulse energy relies on an accurate absolute calibration of the radio antennas at LOFAR. The calibration has been improved with respect to the previous analysis \cite{Mulrey:2019}.
It uses the emission from the Galaxy, which enters the measured time traces of the antenna signals as `noise'.
The galactic emission model LFMap \cite{Polisensky:2007} is used to obtain the contribution of each point on the sky at a given (sidereal) time.
This is integrated over the visible sky, using the antenna response as known from \cite{Schellart:2013}. 
Apart from this, the contributions of electronic noise in multiple stages of the signal chain have been fitted, by comparing the measured variations with sidereal time to the curve from LFMap.

Uncertainties in the calibration, and in the directional dependence in the antenna model, translate into systematic uncertainties on the energy estimates. The \Xmax-estimate is, at least to lowest order, not affected, as all signals arrive from the same direction (to within a degree) for a given air shower. 
Systematic uncertainties on the calibration amount to $\unit[13]{\%}$ in total, where the main contributor (by $\unit[11]{\%}$) is the uncertainty on the sky temperature in our frequency range. The other, minor contributions are uncertainties in the electronic noise levels and from the antenna model.

In \cite{Mulrey:2020}, a cross-check is described between the energy scales determined from radio and from the particle detectors, respectively. For the particle-based reconstruction, one compares the particle footprint from the best-fitting Corsika shower to the signals at the LORA detectors.
The conversion from particles reaching the ground to LORA signals is simulated using the GEANT4 simulation package \cite{Geant4:2003}, a procedure described further in Sect.~\ref{sect:particlebias}.
Agreement within $\unit[10]{\%}$ was found between the resulting energy scales from radio and from particles.

Additional contributions to the systematic uncertainty on primary energy have been tested and were found to be small.
These arise from the choice of simulation code, and from the choice of hadronic interaction model within the simulation code. 
Switching from CoREAS to ZHAireS \cite{ZHAireS:2012} yields a difference below $\unit[3]{\%}$. Thus, importantly, two independent simulation codes aimed at detailed simulations produce very close radio energy levels.
A cross-analysis of the QGSJetII-04 interaction model versus Sibyll-2.3c gives another contribution of $\unit[3]{\%}$.

The resulting systematic uncertainty is $\unit[14]{\%}$, from adding the contributions in quadrature.
This is a considerable improvement from the $\unit[27]{\%}$ in the previous analysis based on the particle detectors.

The statistical uncertainty on the energy estimate follows from our Monte Carlo uncertainty analysis per shower (see Sect.~\ref{sect:method_xmax}).
The average values for the uncertainties on energy and log-energy, i.e.,~$\sigma_E$ and $\sigma_{\log E}$, corresponds to $\unit[9]{\%}$. 
Again, this is a notable improvement over the $\unit[32]{\%}$ uncertainty in \cite{Buitink:2016}, arising from the large number of radio antennas compared to the 20 particle detectors used earlier.

\subsection{Including local atmospheric parameters}
To improve the accuracy of the simulations and the \Xmax reconstructions, we have co-developed an updated version of CORSIKA and CoREAS, which allows to include local atmospheric altitude profiles of density and refractive index into the simulation runs (v7.7100 includes the update). 

The atmospheric parameters at the time of each air shower are taken from the Global Data Assimilation System (GDAS) \cite{GDAS}, which gives pressure, temperature, and humidity in 24 layers in the atmosphere. These are data used, e.g.,~in weather models.

The important quantities for us are the altitude profiles of density and refractive index. The density profile determines the amount of matter traversed by the particles along the shower evolution, and therefore the shower geometry depends on this. 
Due to natural variations in air pressure and temperature, the geometric distance to \Xmax may be under- or overestimated, leading to a systematic error per shower on the order of 15 to $\unit[20]{g/cm^2}$.
In the earlier analysis of \cite{Buitink:2014}, the GDAS density profile was used to correct to first order the \Xmax estimate from simulations using the US Standard Atmosphere.

The refractive index $n$ and its variations are important for the radio emission processes. The refractive index is a function of both the density and the humidity.
Natural variations in $n$ make the Cherenkov angle wider or narrower, thus affecting the intensity footprint on the ground \cite{Corstanje:2017}. 
Typical variations of $(n-1)$ are on the order of $\unit[4]{\%}$, and introduce a systematic error on the inferred \Xmax. From simulations, this error was found to be about 4 to $\unit[11]{g/cm^2}$, depending on the zenith angle.

Residual uncertainties in $(n-1)$ as taken from GDAS temperature, pressure, and humidity are about $\unit[0.5]{\%}$. From this, uncertainties on \Xmax are on the order of 1 to $\unit[2]{g/cm^2}$, and will vary between positive and negative from one shower to another, adding to the statistical uncertainty per shower.
We have thus removed a systematic uncertainty that is important for precision measurements.

In CORSIKA, five layers are used to parametrize the atmospheric density profile as a function of altitude. In each layer (except for the top layer), the density is set to fall off exponentially with altitude, with a scale height as a free parameter.
We have used least-squares curve fitting to determine the optimal parameters to match the five-layer model atmosphere to the GDAS representation \cite{Mitra:2020}. 
The error on \Xmax induced by the five-layer approximation was found to be about $\unit[4]{g/cm^2}$, and adds to the statistical uncertainty per shower. It introduces a systematic uncertainty of 1 to $\unit[2]{g/cm^2}$, depending on altitude, hence taken as $\unit[2]{g/cm^2}$.

\subsection{Bias-free sample selection}\label{sect:biasfree}
In this section we show how to apply fiducial cuts, i.e.,~to reject showers that would introduce a composition bias to the sample.
Cuts are made only based on the simulated ensemble of showers, not on the measured data (for instance, through fit quality).
For the composition measurement, we aim to obtain a sample which is unbiased in \Xmax. We do not expect, however, to obtain a sample reflecting the natural cosmic-ray energy spectrum, as the effective exposure area, both on the ground and on the sky, depends strongly on energy.

A bias may arise from the particle detector trigger, which is reached more easily for showers penetrating deeper into the atmosphere (high \Xmax).
Another, opposite source of bias arises from the radio detection threshold. We require at least 3 LOFAR stations to detect significant pulses for a given shower. Showers with low \Xmax have a larger radio footprint, and hence are more likely to trigger three LOFAR stations.

We analyze each measured shower given its energy, reconstructed shower core position, and incoming direction.
The central requirement is that this shower would have produced a trigger in both the particle detectors and in the radio data, if it had any other value of \Xmax in the natural range. Moreover, it must meet the core reconstruction quality criterion explained below.
As noted in Sect.~\ref{sect:method_xmax}, our simulated ensemble for each measured shower is based on a pre-selection from 600 random showers simulated with CONEX. This sufficiently represents the natural range of \Xmax, as our (final) dataset is smaller than this.

A dataset comprising all measured showers that meet this requirement is then unbiased in \Xmax, so this requirement is a sufficient condition.
Due to the irregular array layout and moderate event count, a per-shower inclusion criterion is more efficient than attempting to construct a fiducial volume in parameter space (which would also be rather irregular).

\subsubsection{Removing selection bias arising from the particle trigger}\label{sect:particlebias}
For each measured shower, we use the set of all simulated showers, including their particle content, to see if each simulated shower would have triggered LORA. 

For this, we use the GEANT4 simulation tool \cite{Geant4:2003}, which simulates the particles traversing the detectors and their deposited energy. The simulation of the LORA detectors was also used in the measurement of the cosmic-ray energy spectrum presented in \cite{Thoudam:2016}.
Only if all showers in the ensemble are able to trigger, the measured shower is included in the sample. 

From CORSIKA we obtain a list of particles reaching the ground, with their respective positions and momenta. In the GEANT4 simulation, this is converted to an energy deposit at the detector locations. We divide the energy deposit by an average value of $\unit[6.2]{MeV}$ per particle.
The value of $\unit[6.2]{MeV}$ arises from the most probable energy deposit of single, high-energy muons from an all-sky distribution \cite{Thoudam:2016}. Although the muons vary in energy, their deposit is nearly constant with energy. 

At the time of each measured shower, we note the trigger threshold of each detector, which was derived during operation from the baseline and standard deviation of its signal time trace. This can be expressed in equivalent muons.
When particles hit a detector and produce enough energy deposit, it will trigger. This is subject to Poisson statistics.
We evaluate the probability of having $\geq n$ particles giving an energy deposit of $\unit[6.2]{MeV}$ each, where $n$ is the first integer above the ratio trigger threshold / 1 equivalent muon.

In our trigger setup, a number $k$ out of 20 LORA detectors must trigger in coincidence for the radio data of the air shower to be recorded. The threshold $k$ has been variable over the years of measurements, where $k=13$ was the most common value. Changes have been made mainly when one or more detectors were down. For each measurement, we use the trigger setting at that time.
Hence, also from all simulated showers we require that with a probability of $\unit[99]{\%}$, at least $k$ detectors would trigger (due to statistical fluctuations, the probability cannot reach exactly $\unit[100]{\%}$).

This test has a tendency to remove showers from the sample which have large reconstructed \Xmax values, i.e.,~at relatively low altitude in the atmosphere, and/or high inclination. In this case, the given measured shower has produced a trigger, but had its \Xmax been lower, the number of particles would have been too small.
Similarly, showers with low energy and/or a core position far from the LORA detectors are more likely to be rejected.

\subsubsection{Removing bias arising from the radio detection threshold}\label{sect:radiobias}
We perform a test against bias from the finite radio detection threshold. 
The criterion is, similar to the particle detection bias test, that the radio signal for each simulated shower in the ensemble would have been detected above the noise in at least three LOFAR stations. 

To this end, we take the core position of the shower that fits best to the LOFAR-measured shower, and position also all other simulated showers here with respect to LOFAR. 
From the best-fitting shower, we have a fitted scale factor relating simulated to measured pulse energy.
Using this scale factor, we obtain the pulse intensities for each simulated shower and for each antenna. 
The noise intensities from the LOFAR-measured showers are taken as reference, and a threshold criterion is set as an energy signal-to-noise ratio of 6 in each antenna. 
In the data processing pipeline, we have a (somewhat arbitrary) threshold requiring half of the antennas per station to trigger to have a `good' detection. Although that detection is amplitude-based, an energy signal-to-noise ratio of 6 was found to be slightly conservative, and otherwise in good agreement with the amplitude threshold detection.
 
This test typically rejects showers with a small reconstructed \Xmax value, i.e.,~relatively high in the atmosphere, and/or zenith angle; a shower with the same parameters would then have a much smaller radio footprint at high \Xmax, which may not be able to trigger three LOFAR stations.

\subsection{Reconstruction quality cuts}
The procedure described in Sect.~\ref{sect:method_xmax} to infer the uncertainty on \Xmax is also useful as a test of the reconstruction quality of the radio signal. Apart from the \Xmax uncertainty, it also gives an uncertainty on the fitted shower core position and on the energy. These uncertainties are calculated from the entire simulated ensemble, and hence they are the same for each simulated shower being tested by the two above procedures. 

\begin{figure}[t]
\begin{center}
\includegraphics[width=0.80\textwidth]{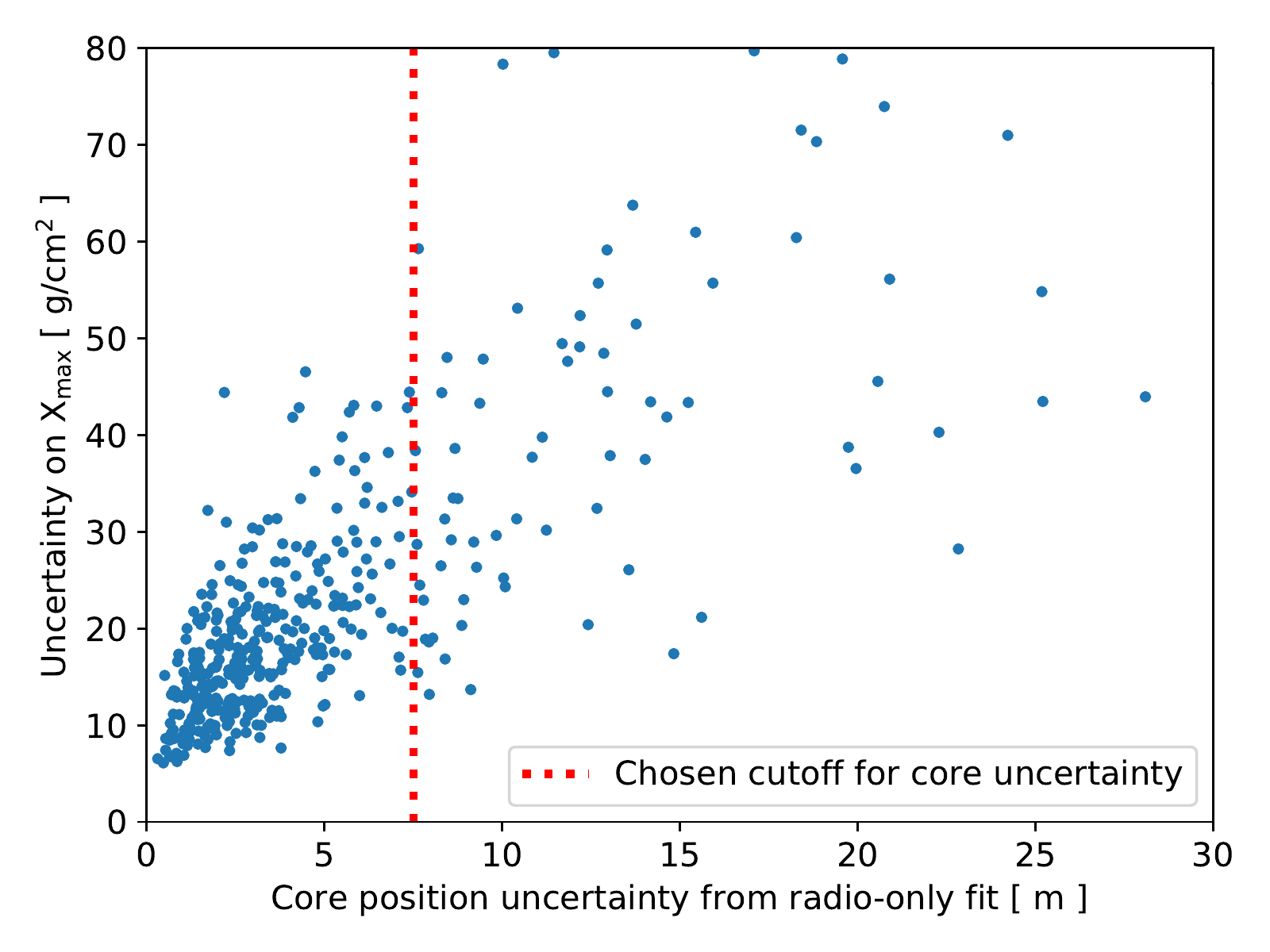}
\caption{The uncertainty on \Xmax versus the uncertainty on the core position, per measured shower, after cuts on energy and fiducial selection criteria. The dotted line indicates the chosen cutoff. }
\label{fig:xmax_core_uncertainty}
\end{center}
\end{figure}

From the three uncertainties, the precision of the core position reconstruction is arguably the most relevant indicator of overall shower reconstruction quality. When this precision is low, one cannot expect either \Xmax or energy to be reconstructed accurately.
Shown in Fig.~\ref{fig:xmax_core_uncertainty} is the uncertainty on \Xmax versus the core position uncertainty. 
They are clearly correlated, and a cut on the reconstruction uncertainty at 7.5 meters was found to be sufficient to reject the majority of poorly reconstructed showers, while retaining showers with low \Xmax uncertainty.

The appearance of poorly reconstructed showers, despite meeting the other criteria, comes mainly from the position of some showers with respect to the LOFAR array geometry. Most notably, when the core position is outside the array and/or only three stations have been triggered at low signal-to-noise ratio, the reconstruction precision becomes well below average.
This criterion catches these cases automatically.

\subsection{Systematic uncertainties}
Our method to determine \Xmax is affected by the following systematic uncertainties, which are summarized in Table~\ref{table:syst_uncertainties}.
The choice of the hadronic interaction model used in CORSIKA, in this case QGSJetII-04, introduces a systematic uncertainty of $\unit[5]{g/cm^2}$ \cite{Buitink:2016} in the \Xmax measurements, due to minor differences in radio footprints when changing the model, for example,~to EPOS-LHC \cite{EPOSLHC:2013}. 
The choice of hadronic interaction model also causes another, larger uncertainty in the composition analysis, as the average \Xmax for a given element varies by up to about $\unit[15]{g/cm^2}$ between models. This is treated separately by repeating the composition analysis with different models.

Residual systematic uncertainties due to variations in the atmosphere, local weather etc.~are about $\unit[2]{g/cm^2}$ from the five-layer approximation of CORSIKA. This approximation also produces an additional statistical uncertainty of $\unit[4]{g/cm^2}$ which is added in quadrature to the statistical uncertainty on \Xmax per shower.
A systematic uncertainty, or bias, in averages of \Xmax may arise from possible residual bias after applying the above selection criteria. 
We test this in Sect.~\ref{sect:residualbias}, obtaining a value of $\unit[3.3]{g/cm^2}$ to be added as a systematic uncertainty on \Xmax.
Hence, a total systematic uncertainty on \Xmax of $\unit[7]{g/cm^2}$ follows. 
This is comparable to the systematic uncertainty on \Xmax in the measurements of \cite{Auger_depth:2014} who find a value between 7 and $\unit[10]{g/cm^2}$ for primary energies above $\unit[10^{17.8}]{eV}$.

\begin{table}[h!]
\caption{Systematic uncertainies in the \Xmax reconstruction}
\centering
\begin{tabular}{l l l}
\hline\hline
 & Syst.~uncertainty & Added stat.~unc.~\\ [0.5ex] 
\hline
Choice of hadronic interaction model & $\unit[5]{g/cm^2}$ &  \\
Remaining atmospheric uncertainty & $\sim \unit[1]{g/cm^2}$ & $\sim \unit[2]{g/cm^2}$ \\
Five-layer atmosphere CORSIKA & $\unit[2]{g/cm^2}$ & $\unit[4]{g/cm^2}$ \\
Possible residual bias & $\unit[3.3]{g/cm^2}$ &  \\ 
Curve fit for $\chi^2$ optimum & $\leq \unit[1]{g/cm^2}$ &  \\
\hline
Total, added in quadrature & $\unit[7]{g/cm^2}$ & \\
\hline
\hline
\end{tabular}
\label{table:syst_uncertainties}
\end{table}

When performing the parabolic fit to the $\chi^2$ values per simulation, as in the middle panel of Fig.~\ref{fig:event_fit_example}, a systematic error of up to $\unit[5]{g/cm^2}$ may arise if the fit optimum is not contained in the dense region of simulations. This is removed by simulating extra showers around the optimum when needed. A Monte Carlo simulation shows no residual systematic error ($\leq \unit[1]{g/cm^2}$) if the dense region is positioned asymmetrically around the optimum but does contain it.

The systematic uncertainty in the energy estimate from the radio antennas was found to be $\unit[14]{\%}$, or $0.057$ in $\lg(E)$ \cite{Mulrey:2019}; by convention we write $\lg E \equiv \log_{10} E$.

\section{Mass composition analysis}\label{sect:composition}
Having established the set of showers for the mass composition analysis, we perform statistical analysis on the measured data, being (\Xmax, $\sigma_{X_{\rm max}}$, $\lg E$, $\sigma_{\lg E}$) for each shower.
We make use of the probability density functions of \Xmax as a function of energy and atomic mass number $A$, as parametrized by  \cite{DeDomenico:2013} and updated by \cite{Petrera:2020}. The parametrizations follow a generalized Gumbel distribution, which is a function with 3 parameters, yielding a variable mean, spread, and tail-end asymmetry, respectively. 
The function has been fitted to a large sample of CONEX showers and has a precision within $\unit[2]{g/cm^2}$ for both average and standard deviation of \Xmax, as well as a close fit to the distribution itself; the high-end tail, which drops exponentially, was shown to be well represented. It should be noted that CONEX is a faster but less complete shower simulation method than CORSIKA. Average \Xmax values were found to deviate by 4 to $\unit[5]{g/cm^2}$. This is therefore treated as an additional systematic uncertainty on \Xmax, which for the composition analysis then amounts to $\unit[8]{g/cm^2}$.

\begin{figure}[t]
\begin{center}
\includegraphics[width=0.70\textwidth]{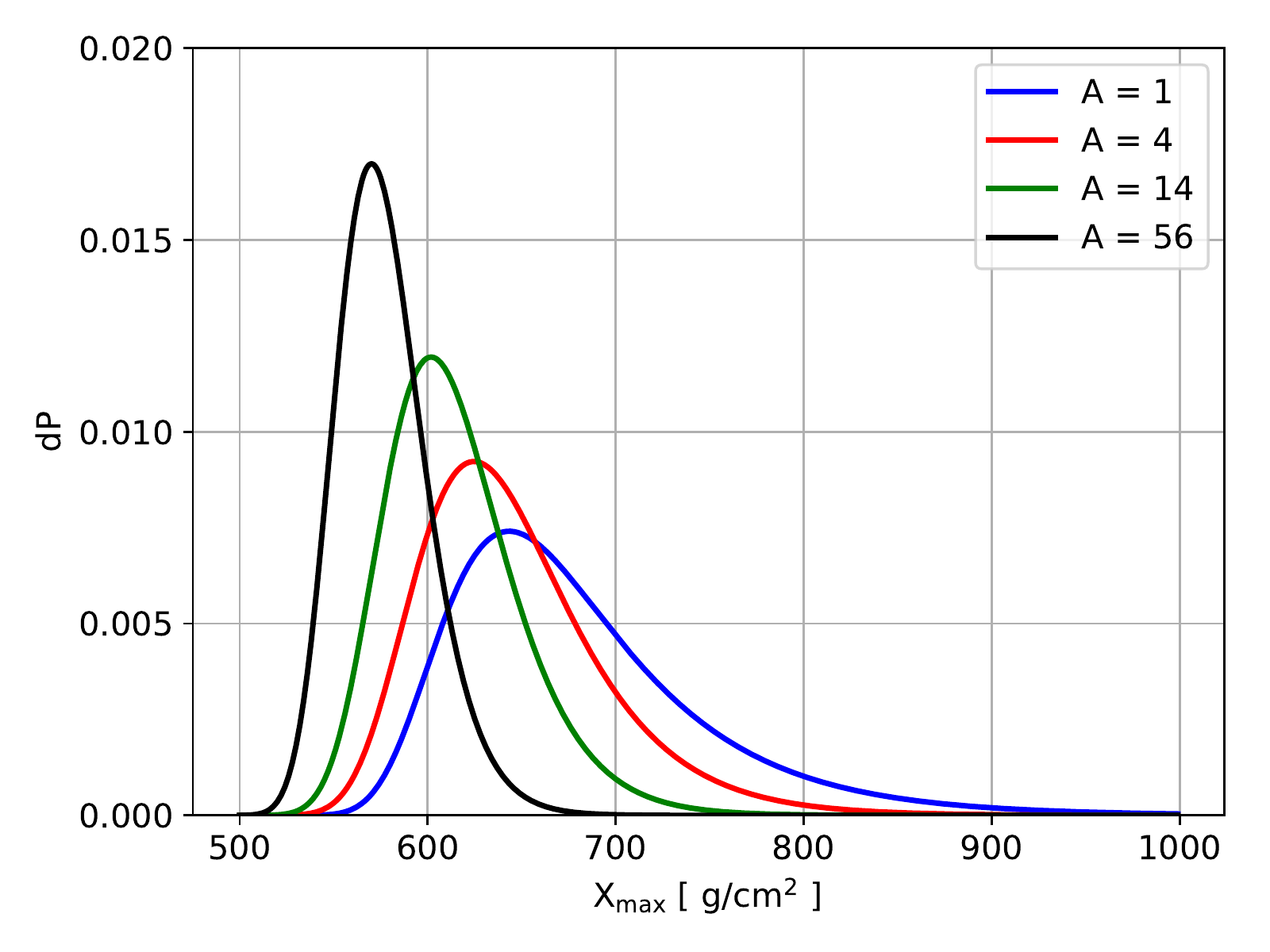}
\caption{The probability density functions for the depth of shower maximum \Xmax, for the elements H, He, N, and Fe, at energy $\unit[10^{17}]{eV}$ and hadronic interaction model QGSJetII-04. }
\label{fig:xmax_gumbel}
\end{center}
\end{figure}

Example curves are shown in Fig.~\ref{fig:xmax_gumbel} for $E=\unit[10^{17}]{eV}$ for protons, helium, nitrogen, and iron nuclei. 
The functions overlap substantially, limiting the extent to which the individual elements can be distinguished. This is the statistical challenge in performing a composition analysis on \Xmax data.
The mean \Xmax shifts approximately proportionally to $\ln A$. Therefore, for a 4-component model of astrophysically relevant elements, a reasonable choice is to take p, He, C/N/O, and Fe. These are roughly equally spaced in $\ln A$; as C, N, and O cannot be readily distinguished, either of them can be chosen as a proxy for all three. We choose nitrogen, as this is in between carbon and oxygen, and the best choice in terms of equal spacing in $\ln A$.

\subsection{Statistical analysis}\label{sect:stat_analysis}
We use an unbinned maximum likelihood method to determine the best-fitting parameters for the four-component composition model. 
This has the advantage of treating each shower separately, instead of relying on \Xmax-histograms and/or binning in energy.
This is especially suitable when the dataset is relatively small and a narrow binning in shower energies is inappropriate.

Given a measured shower with parameters (\Xmax, $\sigma_{X_{\rm max}}$, $\lg E$, $\sigma_{\lg E}$), its likelihood function for a given element is described by the curves in Fig.~\ref{fig:xmax_gumbel}, convolved with a Gaussian for the uncertainties $\sigma_{X_{\rm max}}$ and $\sigma_{\lg E}$. For a mixed composition, the likelihood function is a weighted average of these; the mix fractions maximizing the likelihood is taken as the best-fitting composition.

With this method, a complementary goodness-of-fit test is needed, for which we use a Kolmogorov-Smirnov test, enhanced with Monte Carlo simulation.
This is a simple, well-known method, comparing the cumulative distribution function (cdf) of the best-fit model to the empirical cumulative distribution of the data.
The Kolmogorov-Smirnov test statistic is defined as the maximum difference between the model's cdf $F(X)$ and the empirical distribution $E(X)$:
\begin{equation}\label{eq:ks-test}
K=\sup_X \left| F(X) - E(X) \right|.
\end{equation}
For our case, where the best-fitting distribution has been estimated from data, standard critical values of the test statistic $K$ do not apply. 
Instead, to determine the $p$-value corresponding to $K$, we use parametric bootstrap sampling from $F(X)$, counting how often the $K$-value is larger than the one for the dataset. This tests the null hypothesis that the dataset is a random drawing from $F(X)$. 

The best-fit model \Xmax distribution $F(X)$ is taken as the cumulative integral of the linear combination of \Xmax-distributions $f(\Xmaxmath, E)$ for the best-fitting composition:
\begin{equation}\label{eq:sumdistrib}
f_{\rm sum}(\Xmaxmath) = \frac{1}{N}\sum_{k=1}^N\, \sum_i \, \alpha_i\,f_i(\Xmaxmath,\, E_k)*\mathcal{N}_X(\Xmaxmath,\,\sigma_{X_{\rm max},k}^2) * \mathcal{N}_{\lg E}(\lg E_k,\, \sigma_{\lg E,k}^2),
\end{equation}
summing over all showers (index $k$) and over the elements in the composition model (index $i$). Here, $*$ denotes convolution, in this case with Gaussians corresponding to uncertainties in \Xmax and log-energy.

For the uncertainty analysis we use a likelihood ratio test. 
Denoting the likelihood of the best-fitting composition as $L(\{\hat{\alpha}\})$, we fix one of the element fractions, say the proton fraction, scanning over the range from $0$ to $1$. We then find the maximum likelihood composition given the fixed proton fraction, $L(\alpha_p, \{\hat{\alpha}_i\})$, again optimizing over the free parameters indexed by $i$.
This gives the test statistic $D$:
\begin{equation}\label{eq:likelihood_ratio}
D = 2\, \ln \left( \frac{L(\{\hat{\alpha}\})}{L(\alpha_p, \{\hat{\alpha}_i\})}\right).
\end{equation}
This is nonnegative by definition, and in the large-$N$ limit it follows a chi-squared distribution with 1 degree of freedom, when fixing 1 parameter.
The confidence intervals at significance level $1-p$ then follow directly from the critical values of the chi-squared distribution. 
For confidence levels of $68$, $95$, and $\unit{99}{\%}$, these are $1.00$, $3.84$, and $6.64$, respectively. 

Confidence intervals for two elements simultaneously, such as used in the contour plot Fig.~\ref{fig:contour_p_He} in Sect.~\ref{sect:results_composition} are computed analogously, fixing two parameters instead of one, and noting that the test statistic $D$ then follows a $\chi^2(2)$-distribution.

When splitting the sample into two equal-sized bins, such as done in Sect.~\ref{sect:twobins}, one can use another likelihood ratio test to assess the significance of the difference between results in the two bins.
For instance, in a model with three independent parameters such as used in this analysis, splitting into two bins adds three parameters to the (model) description of the data. As a result, the combined (log)likelihood, and test statistic $D$, will be higher than in a single-bin analysis. The difference follows a $\chi^2(3)$-distribution under the null hypothesis that the data in both bins is a drawing from the same model distributions.
This yields a p-value for the difference between the bin results. 

\section{Results: the measured \Xmax distribution}\label{sect:results_xmax}

In the following sections, we present the results regarding statistics on \Xmax, such as the estimate of the mean and standard deviation of the \Xmax distribution. 
After this, the implications for the cosmic-ray composition in our energy range are given, based on the statistical analysis presented in Sect.~\ref{sect:composition}.

The results are based on a dataset of $N=334$ cosmic rays with energies between $10^{16.8}$ and $\unit[10^{18.3}]{eV}$, which pass all selection criteria for a bias-free sample with accurately reconstructable showers, as explained in Sect.~\ref{sect:biasfree}. It is a subset of 720 showers measured in at least three LOFAR stations, of which 469 have a core reconstruction precision better than $\unit[7.5]{m}$. Another 135 showers did not meet the sample selection criteria; their inclusion would lead to a dataset biased in \Xmax. 
The uncertainty on the \Xmax measurement per shower is on average $\unit[19]{g/cm^2}$.
The average fit quality of the best-fitting simulation to the measured LOFAR data is $\chi^2 / {\mathrm{dof}} = 1.19$, indicating a good fit.

\subsection{Mean and standard deviation of \Xmax as a function of primary energy}
We have divided the dataset into energy bins of width 0.25 in $\lg(E / \mathrm{eV})$, and computed the mean and standard deviation in each bin. 

\begin{figure}[h]
\begin{center}
\includegraphics[width=0.70\textwidth]{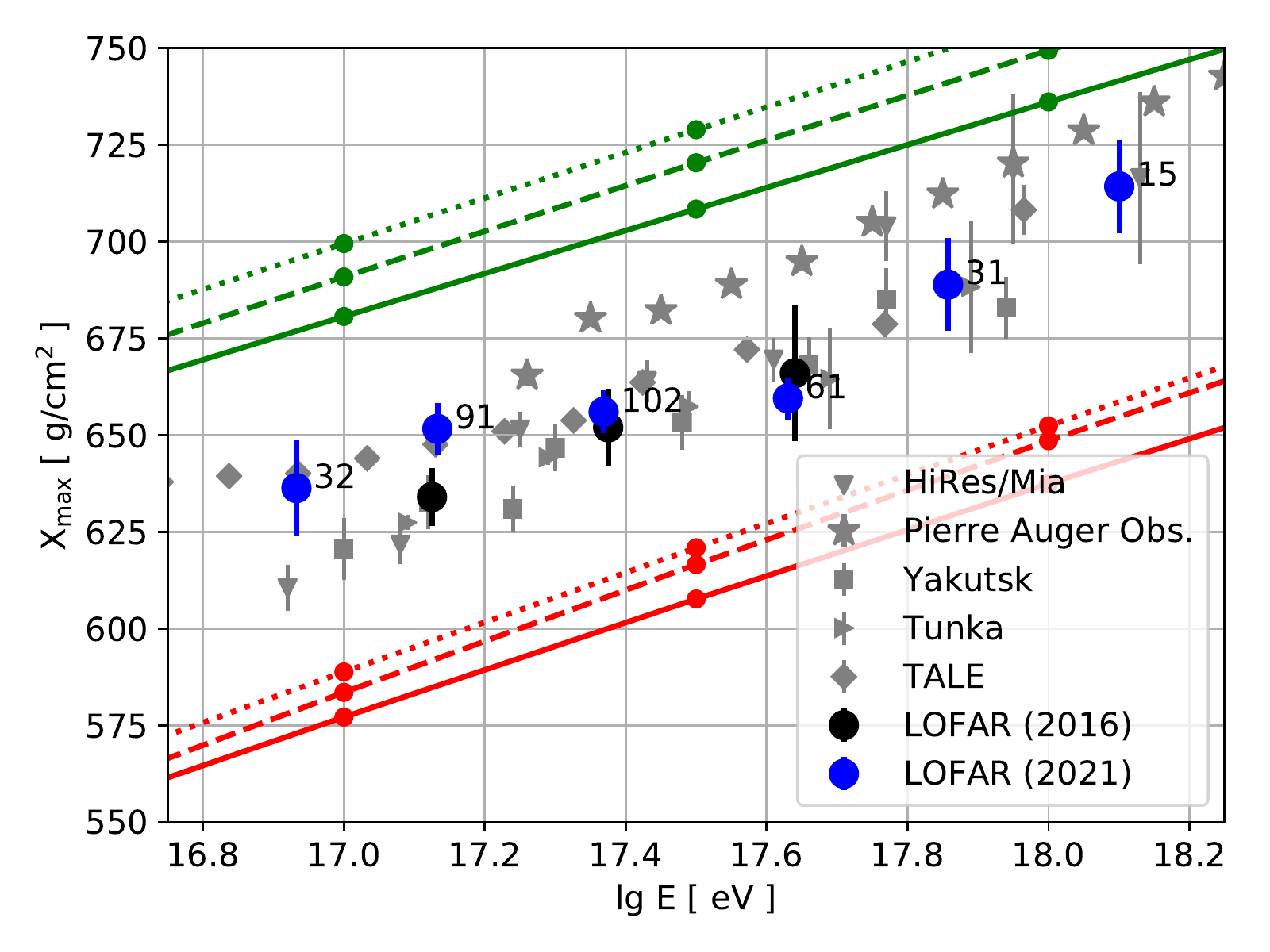}
\caption{The average depth of shower maximum \Xmax, as a function of primary particle energy. The annotated numbers indicate the number of showers in each bin, and the error margins indicate the uncertainty on the mean of the \Xmax distribution. The upper lines indicate the mean values expected for protons, from simulations with QGSJetII-04 (solid), EPOS-LHC (dashed) and Sibyll-2.3d (dotted). The lower lines show the mean predicted values for iron nuclei. For comparison, results from Pierre Auger \cite{Yushkov:2019}, Yakutsk \cite{Knurenko:2015}, Tunka \cite{Prosin:2015}, HiRes/Mia \cite{Sokolsky:2011}, and TALE \cite{TALE:2020} are included.} 
\label{fig:avg_xmax}
\end{center}
\end{figure}

The sample averages are shown in Fig.~\ref{fig:avg_xmax}. The given uncertainty is the uncertainty on the mean of the \Xmax distribution, i.e.,~$\sigma / \sqrt{N_{\rm bin}}$, with sample standard deviation $\sigma$. 
For positioning the points we have used the average log-energy inside each bin.
Two showers above $\lg (E/\mathrm{eV}) = 18.25$ were discarded, as no meaningful average can be taken from them. 

For comparison, results are included from the Pierre Auger Observatory \cite{Yushkov:2019}, HiRes \cite{Sokolsky:2011}, Tunka \cite{Prosin:2015}, and Yakutsk \cite{Knurenko:2015}. We also include recent results from TALE \cite{TALE:2020}, noting that their method to infer a bias-corrected $\left<\Xmaxmath\right>$ is different and assumes the EPOS-LHC hadronic interaction model.

\begin{figure}[h]
\begin{center}
\includegraphics[width=0.70\textwidth]{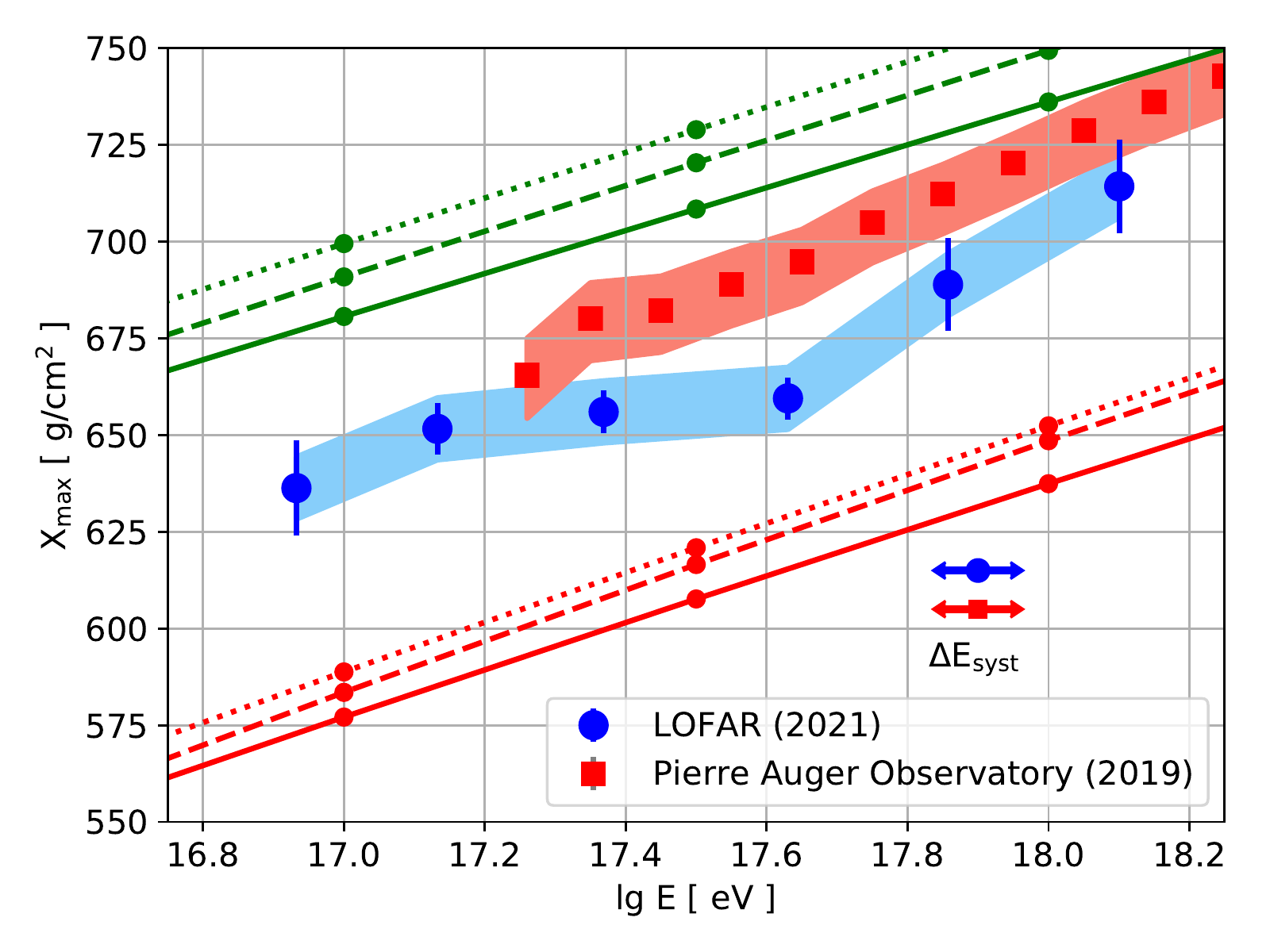}
\caption{Average \Xmax versus primary energy, for LOFAR and Pierre Auger Observatory, with colored bands indicating their systematic uncertainty on \Xmax. The uncertainty margins per data point are statistical uncertainties only. The systematic uncertainty on energy is the same for both experiments, and is indicated by the arrows to the lower right.}
\label{fig:avg_xmax_systematics}
\end{center}
\end{figure}

\begin{figure}[t]
\begin{center}
\includegraphics[width=0.70\textwidth]{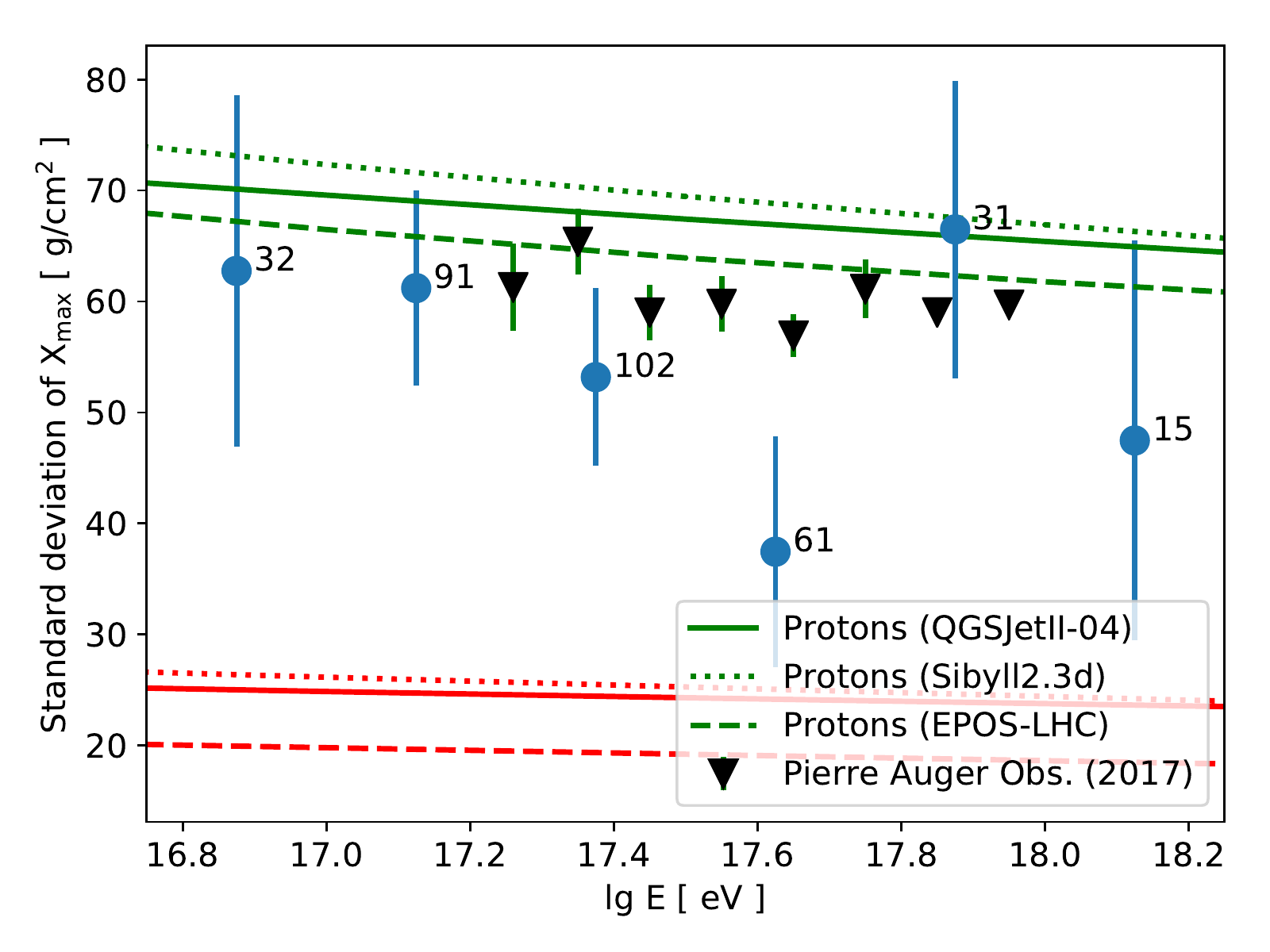}
\caption{The standard deviation of \Xmax as a function of primary particle energy. The margins indicate the uncertainty on the standard deviation of the \Xmax distribution. Results from the Pierre Auger Observatory are shown, together with the values from simulations of protons and iron nuclei (high and low lines, respectively). }
\label{fig:stddev_xmax}
\end{center}
\end{figure}
\begin{figure}[h]
\begin{center}
\includegraphics[width=0.80\textwidth]{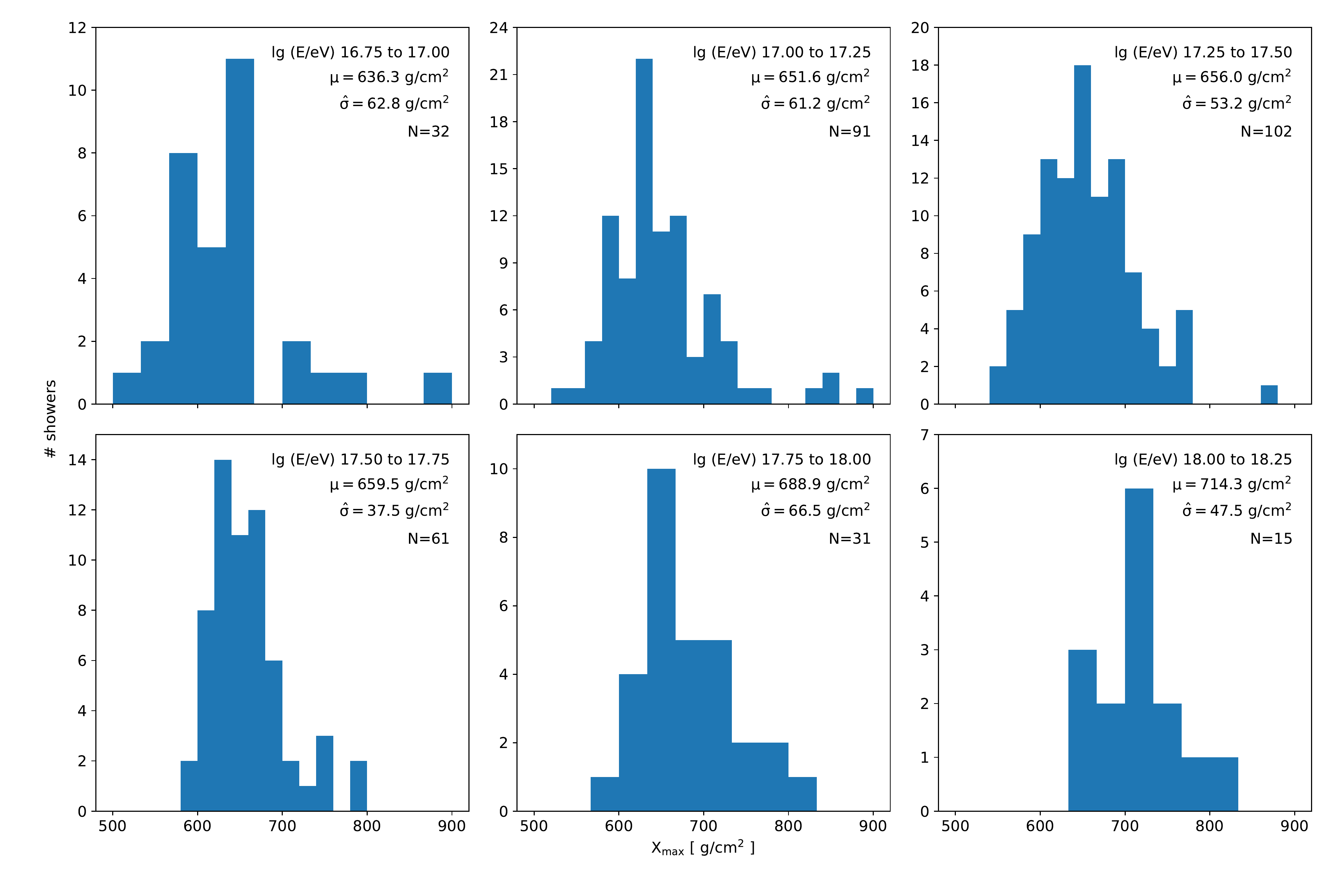}
\caption{Histograms of \Xmax for each energy bin. The average ($\mu$) and standard deviation estimates ($\hat{\sigma}$) per bin are as shown in Figs.~\ref{fig:avg_xmax} and \ref{fig:stddev_xmax}, respectively. } 
\label{fig:histograms_per_bin}
\end{center}
\end{figure}

The differences with respect to the earlier LOFAR results \cite{Buitink:2016} can be explained through statistical fluctuations, and from the revised treatment of systematic effects including the atmosphere and the radio-derived energy scale. The lowest-energy data point stands out somewhat, with a difference of $\unit[17.6]{g/cm^2}$, at statistical uncertainties of 7.5~and~$\unit[6.8]{g/cm^2}$ for the older and newer result, respectively.
Such a difference in one of three overlapping data points is not unreasonable just from statistics; it is also possible that improvements in fiducial selection criteria make some difference here, as differences are expected to appear especially at lower energies where signals are closer to trigger thresholds.

The average \Xmax agrees reasonably well with the other experiments such as Tunka, Yakutsk, HiRes/Mia, and TALE, especially for $\lg E > 17.2$.
However, the results from the Pierre Auger Observatory, which is the largest experiment, are somewhat higher starting at the bin around $\lg E = 17.325$. Their statistical uncertainty is smaller than the plotted symbols, arising from a high number of showers (1000 to 2600) per individual bin.
Systematic uncertainties on \Xmax in this energy range are about $\unit[11]{g/cm^2}$ for Auger \cite{Bellido:2017}, and about $\unit[7]{g/cm^2}$ for LOFAR. Additionally, there is a systematic uncertainty in energy, which for LOFAR as well as Auger \cite{Dawson:2019} is about 0.057 in $\lg E$. 

To better compare the results of LOFAR and Pierre Auger Observatory and their systematic uncertainties, we have plotted these separately in Fig.~\ref{fig:avg_xmax_systematics}. The band plots are seen to have little to no overlap, although systematic uncertainties in energy could shift either result horizontally in this plot, according to the margins indicated by arrows. 

Thus, there is tension between these results. There is a notable difference in methodology to measure \Xmax, being direct fluorescence detection versus radio detection with Corsika/CoREAS simulations. In both cases, considerable attention was given to estimating systematic uncertainties from different contributions.
In the range where LOFAR and Auger overlap, our results are in agreement with the other shown experiments on the Northern hemisphere, including the recent results from TALE. 
Also, our earlier results from 2016 are consistent with the present results, while for the latter, systematic uncertainties on energy and atmospheric effects have been lowered considerably. 
Hence, the apparent difference is not fully explained at present.

In Fig.~\ref{fig:stddev_xmax}, we show the standard deviation in each bin, along with its uncertainty. 
To calculate these, as an estimator $\hat{\sigma}$ of the underlying \Xmax-distribution's standard deviation, we subtract the variance caused by the \Xmax uncertainty per measured shower:
\begin{equation}
\hat{\sigma} = \sqrt{\sigma^2 - \frac{1}{N_j}\sum_{i=1}^{N_j} u_i^2},
\end{equation}
with $\sigma$ the sample standard deviation, $u_i$ the \Xmax uncertainty on each shower, and $N_j$ is the number of showers in energy bin $j$.
The uncertainty on the standard deviation of the distribution is estimated using a parametric bootstrap, taking the best-fit mass composition from Sect.~\ref{sect:results_qgsjet}. In this method one needs to assume a particular model distribution, but it is suitable for estimating uncertainties in small samples. Switching the hadronic interaction model was found to make little difference here.

The results are consistent with those from the Pierre Auger Observatory, except for one bin around $\lg E = 17.625$.
However, a caveat is the relatively low number of showers per bin, and the exponential tail of the \Xmax-distributions.
The showers at the high end of \Xmax, roughly $\Xmaxmath > \unit[800]{g/cm^2}$, appear in our dataset only at low-number statistics level, while their presence shifts the standard deviation considerably upward. In the energy bin from $\lg E = 17.5$ to $17.75$, there happen to be none, thus lowering the sample standard deviation.

The results are summarized in Fig.~\ref{fig:histograms_per_bin}, showing the histograms in each energy bin.

\subsection{Tests for residual bias}\label{sect:residualbias}
In our energy range, the average \Xmax of the measured showers is expected to be independent of shower parameters such as the zenith angle. This follows from the fact that the cosmic-ray composition in our energy range is independent of time and incoming direction, as far as is known from experiments.
This allows to perform a test of our sample for a residual bias in \Xmax due to variations in these parameters.
Our sample of 334 showers has an average \Xmax of $659 \pm \unit[3.3]{g/cm^2}$.
\begin{figure}[b]
\begin{center}
\includegraphics[width=0.7\textwidth]{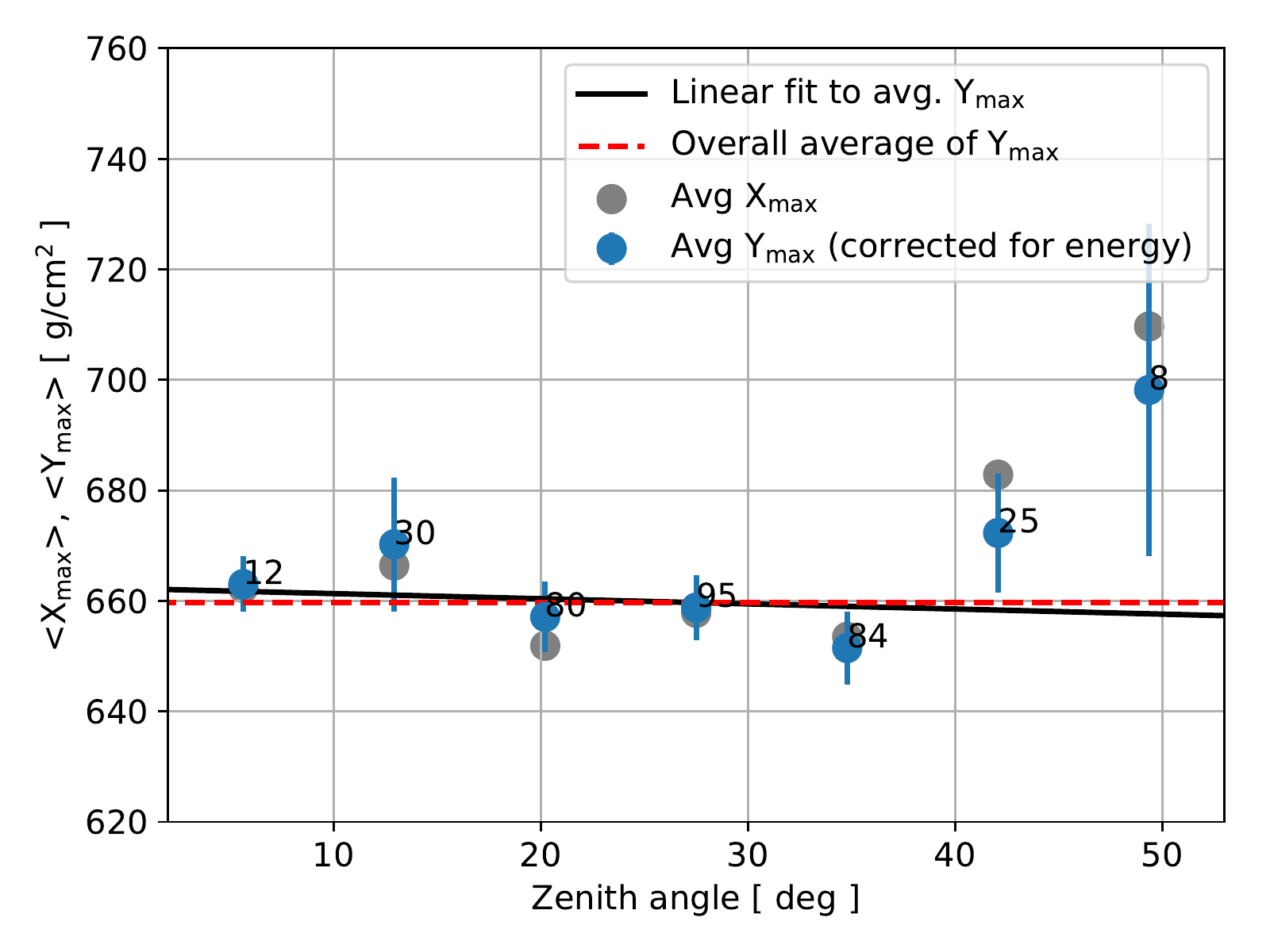}
\caption{Average $Y$ (from Eq.~\ref{eq:xmax_energy_shift}) versus zenith angle, together with a constant and linear fit. }
\label{fig:lgE_xmax_vs_zenith}
\end{center}
\end{figure}
\begin{figure}[h]
\begin{center}
\includegraphics[width=0.75\textwidth]{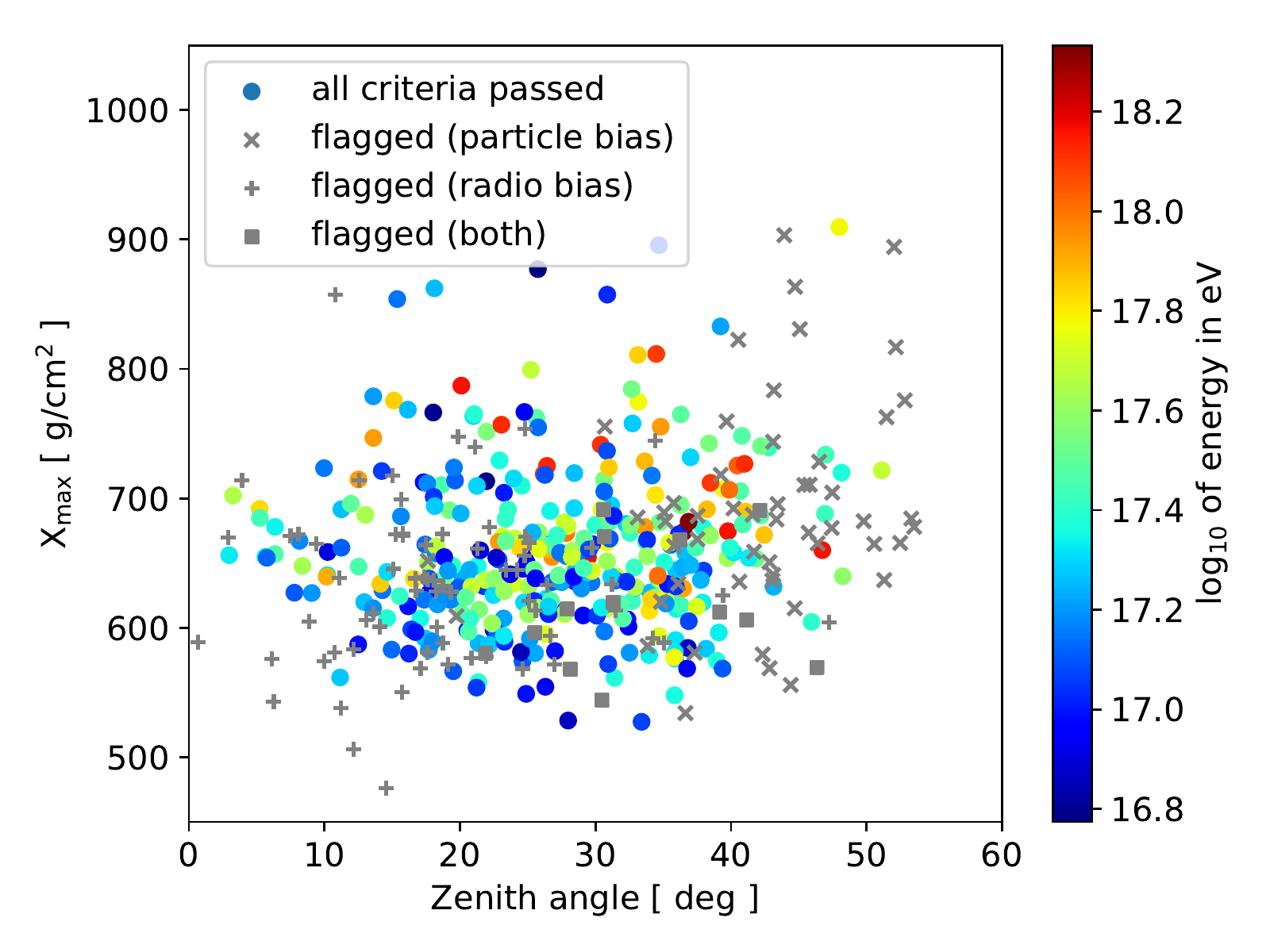}
\caption{Scatter plot of \Xmax versus zenith angle, for all $469$ showers with core reconstruction precision better than $\unit[7.5]{m}$. Colored circles represent the 334 showers passing all criteria, with the color denoting their energy. The showers flagged by the particle and radio bias tests are also shown.}
\label{fig:xmax_scatter_zenith}
\end{center}
\end{figure}

As discussed in Sect.~\ref{sect:biasfree}, a biased sample would readily show a dependence of the average \Xmax on zenith angle. 
However, the average \Xmax also depends on the energy; its expected value is to good approximation linear in $\lg E$ over our energy range. 
From the parametrization using Gumbel distributions, as discussed in Sect.~\ref{sect:composition}, and for the QGSJetII-04 hadronic interaction model, we find, for a factor 10 increase in energy, an average rise in \Xmax (elongation rate) of 55.4, 57.8 and $\unit[60.3]{g/cm^2}$, for protons, nitrogen, and iron nuclei, respectively. This is in good agreement with the elongation rate of $\unit[58]{g/cm^2}$ predicted, for example, by the Heitler-Matthews model \cite{Matthews:2005}.

The average (log-)energy of our measured showers tends to rise with zenith angle. This is no problem, as long as there is no bias in \Xmax.
A possible residual bias in \Xmax, corrected for the influence of varying energy, is evaluated by introducing a parameter $Y$ for each shower, as
\begin{equation}\label{eq:xmax_energy_shift}
Y = \Xmaxmath + 57\,(\lg\,(E/\mathrm{eV}) - 17.4)\;\unit{g/cm^2},
\end{equation}
where 17.4 is approximately the average value of log-energy in our sample.

The results are shown in the right panel of Fig.~\ref{fig:lgE_xmax_vs_zenith}, together with a linear fit. 
The uncertainty margins are once again given by the standard error of the mean.
A constant fit of $Y=660$ as well as a linear fit are shown. 

The linear fit has a slope parameter of $-0.10 \pm 0.30$. Hence, the slope is compatible with zero, and no residual bias is evident. The high value near 700 for the rightmost bin appears suggestive, but as it contains only eight showers and has a correspondingly large uncertainty, it is not significant.
The constant fit has an uncertainty of $\unit[3.3]{g/cm^2}$. A bias at this level cannot be ruled out, hence this is added as a contribution to the systematic uncertainty on \Xmax.

We also show a complete scatter plot of \Xmax versus zenith angle, in Fig.~\ref{fig:xmax_scatter_zenith}.
This plot shows the effect of the bias tests for the radio and particle detectors and the corresponding fiducial cuts (Sect.~\ref{sect:particlebias} and \ref{sect:radiobias}).
As expected, the particle bias test flags most events at high inclination and high \Xmax, especially above 45 degrees. 
The radio bias test flags mostly the opposite region, low \Xmax and low inclination.

Consequently, we see only few showers passing the tests at $\theta < 10^\circ$, and there are only 8 in the highest zenith angle bin above $45 ^\circ$. 
The plot makes clear that the fiducial cuts from Sect.~\ref{sect:biasfree} are necessary, as there would have been a strong zenith angle dependence, and thus a biased \Xmax-sample, had it been omitted.

\clearpage

\section{Mass composition results}\label{sect:results_composition}
We have applied the statistical analysis in Sect.~\ref{sect:composition} to the set of 334 showers passing all selection criteria. 
Results are shown in detail for the QGSJetII-04 model, and summarized for the other two considered, which are EPOS-LHC \cite{EPOSLHC:2013} and Sibyll-2.3d \cite{Sibyll:2020}.
We show the results for the full range, followed by a division in two equally-sized energy bins.
Owing to the strong overlap in \Xmax-distributions (see Fig.~\ref{fig:xmax_gumbel}), an $N=334$ dataset is modest-sized for mass composition analysis based on a multi-element model, hence further binning is not appropriate.
The statistical method presented in Sect.~\ref{sect:stat_analysis} is well suited for analysis of a wide energy range without loss of measurement information.

\subsection{Statistics for the QGSJetII-04 hadronic interaction model}\label{sect:results_qgsjet}
\begin{figure}[h]
\begin{center}
\includegraphics[width=0.7\textwidth]{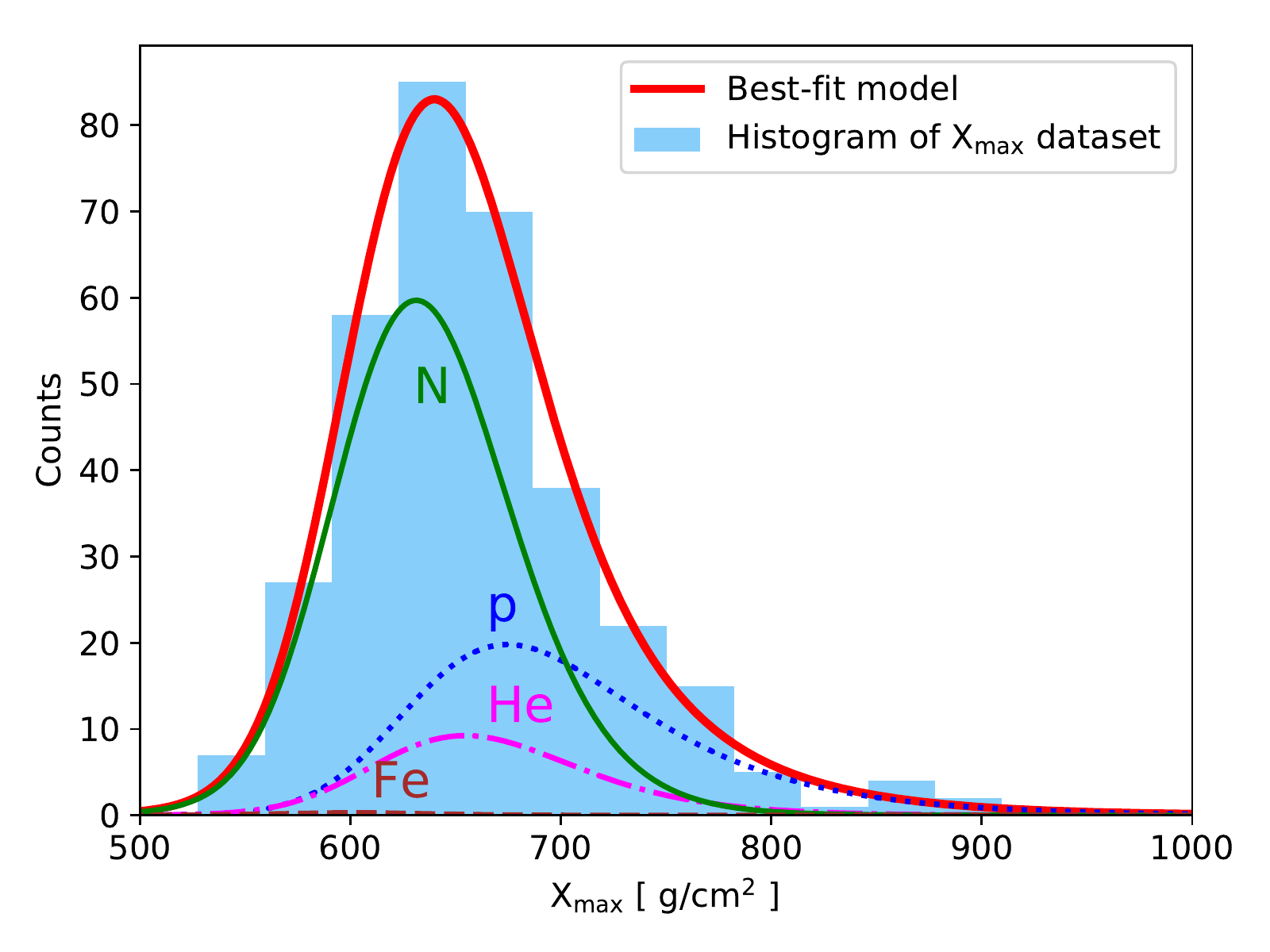}
\includegraphics[width=0.7\textwidth]{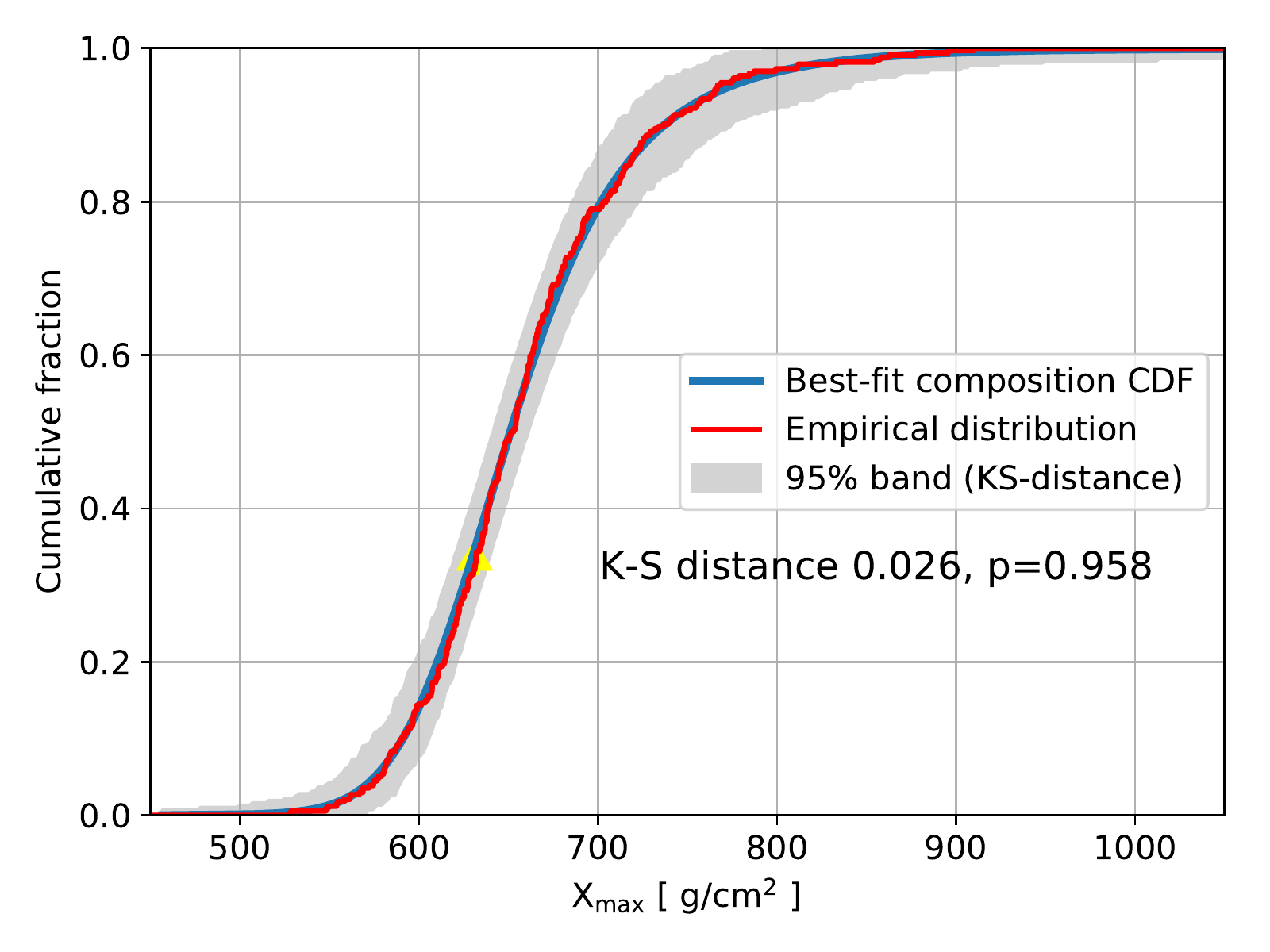}
\caption{Top: A histogram of \Xmax, together with the best-fitting distribution from Eq.~\ref{eq:sumdistrib}. 
Bottom: The cumulative distribution corresponding to the best-fit composition model, together with the empirical distribution from our dataset. The yellow arrow indicates where the distance between CDF and empirical distribution is maximal. The grey band is the envelope of all simulated empirical distributions that have a K-S distance to the CDF below its 95-percentile level. Both graphs indicate a good fit.}
\label{fig:KStest}
\end{center}
\end{figure}

The maximum likelihood estimate was found to be $\unit[28]{\%}$ protons, $\unit[11]{\%}$ helium, $\unit[60]{\%}$ nitrogen and $\unit[1]{\%}$ iron. 
A histogram of \Xmax is shown in Fig.~\ref{fig:KStest} (top), for the full energy range. 
The coverage of this energy range can be summarized by a mean log-energy of $\lg (E/\mathrm{eV}) = 17.39$ and a standard deviation of $0.32$. Hence, the `center of mass' of the dataset lies between $17.07$ and $17.71$. 

The red (solid) curve is the best-fitting distribution, found using the maximum likelihood method and Eq.~\ref{eq:sumdistrib}. The distributions for the elements that make up the best-fitting distribution are also shown, scaled by their respective mix fractions. 

We have tested the goodness-of-fit of the best-fitting model to our dataset, using the Kolmogorov-Smirnov distance between the cumulative and empirical distributions (Sect.~\ref{sect:stat_analysis}) as compared to random drawings from the best-fit distribution.
As shown in Fig.~\ref{fig:KStest} (bottom), the model is a good fit to the data ($p=0.96$). 
Switching the hadronic interaction model to EPOS-LHC or Sibyll-2.3d produces about equally good fits to the data, at $p=0.93$ and $p=0.90$ respectively. Hence, all three models fit the data well, at their respective best-fitting composition. It is of course possible that this would change with a larger dataset and/or smaller energy bins.

We observe that protons and helium are to a significant degree interchangeable in the statistical analysis of our model, given our dataset. 
This is readily seen in the contour plot in Fig.~\ref{fig:contour_p_He}, showing the $D$-statistic for the likelihood ratio test, versus proton and helium fractions. The contours show the allowed regions with confidence levels one-sigma ($\unit[68]{\%}$), $\unit[95]{\%}$, and $\unit[99]{\%}$, respectively. 
Within the one-sigma region, one can exchange helium for protons in a ratio of about 3 to 1. 
The contour plot underlines, for example, that the dataset only allows a very low proton fraction if the helium fraction is rather large instead. 

\subsection{Accounting for systematic uncertainties}\label{sect:systematics}
The systematic uncertainty in \Xmax amounts to $\pm \unit[8.1]{g/cm^2}$ (before roundoff), including the contribution from the CONEX-based parametrizations. The energy uncertainty of $\unit[14]{\%}$, or 0.057 in $\lg E$ has, to first order, the effect of an overall shift of \Xmax in the \Xmax-distributions (see Eq.~\ref{eq:xmax_energy_shift}), of $\unit[3.1]{g/cm^2}$. 
By adding both uncertainties in quadrature, we obtain a systematic uncertainty of $\unit[9]{g/cm^2}$.

Evaluating the composition results for $\Xmaxmath \pm \unit[9]{g/cm^2}$ for all showers, we obtain limits for the best fit, as well as the (expanded) confidence intervals that arise for a systematic shift in this range.
Noteworthy is for example that when the average \Xmax is shifted downward, the helium fraction is fitted much higher at the expense of the nitrogen fraction.
Helium is then favored over nitrogen in the fit, due to the lower expected \Xmax at lower energies, and the longer `tail' of the \Xmax distribution for helium.

\clearpage

\subsection{Results for three hadronic interaction models}\label{sect:results_3models}
\begin{figure}[t]
\begin{center}
\includegraphics[width=0.70\textwidth]{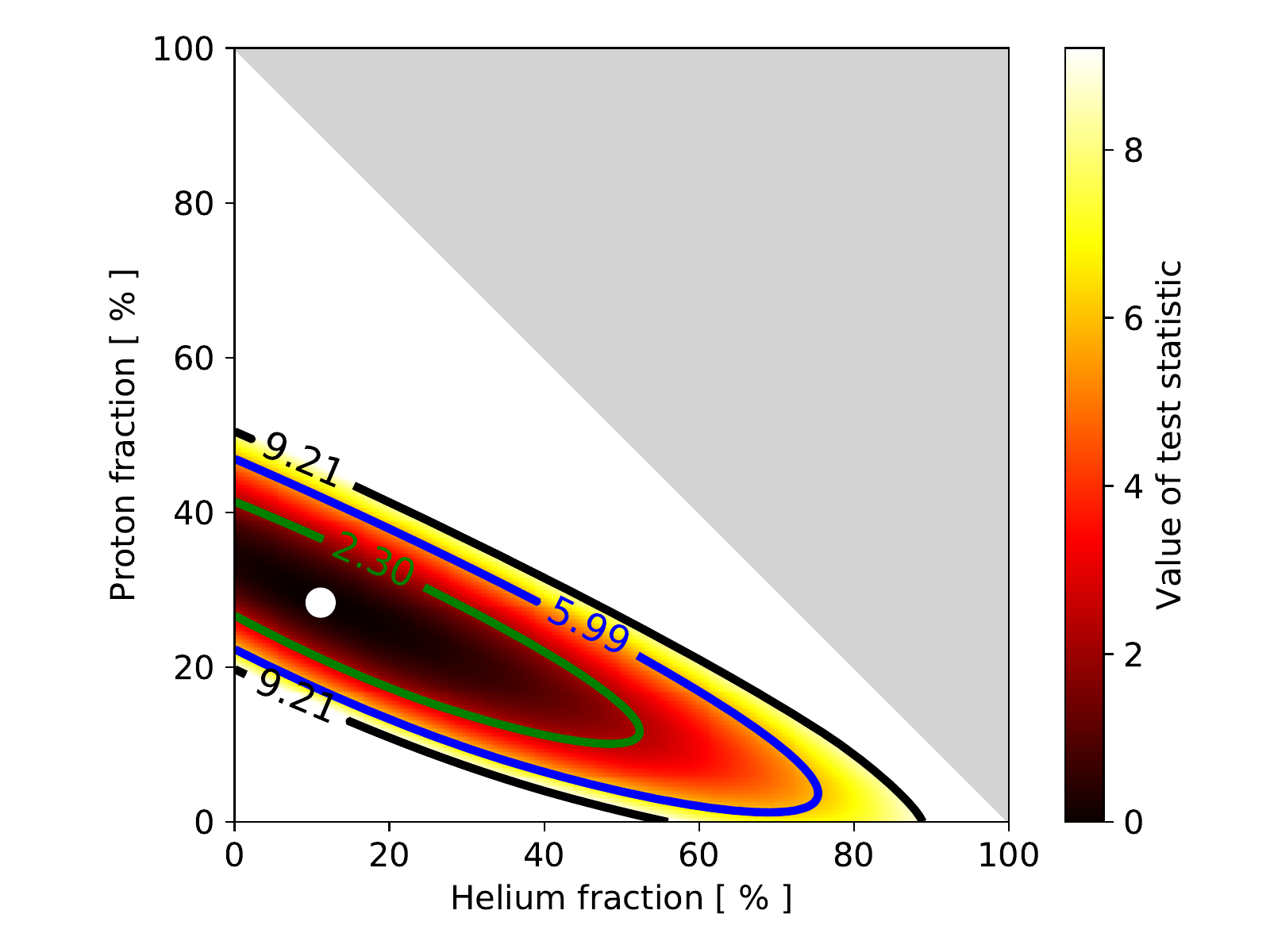}
\caption{Contour plot of the proton and helium fraction, giving the regions consistent with a one-sigma, $\unit[95]{\%}$ and $\unit[99]{\%}$ confidence level, respectively.}
\label{fig:contour_p_He}
\end{center}
\end{figure}

\begin{figure}[h]
\begin{center}
\includegraphics[width=0.7\textwidth]{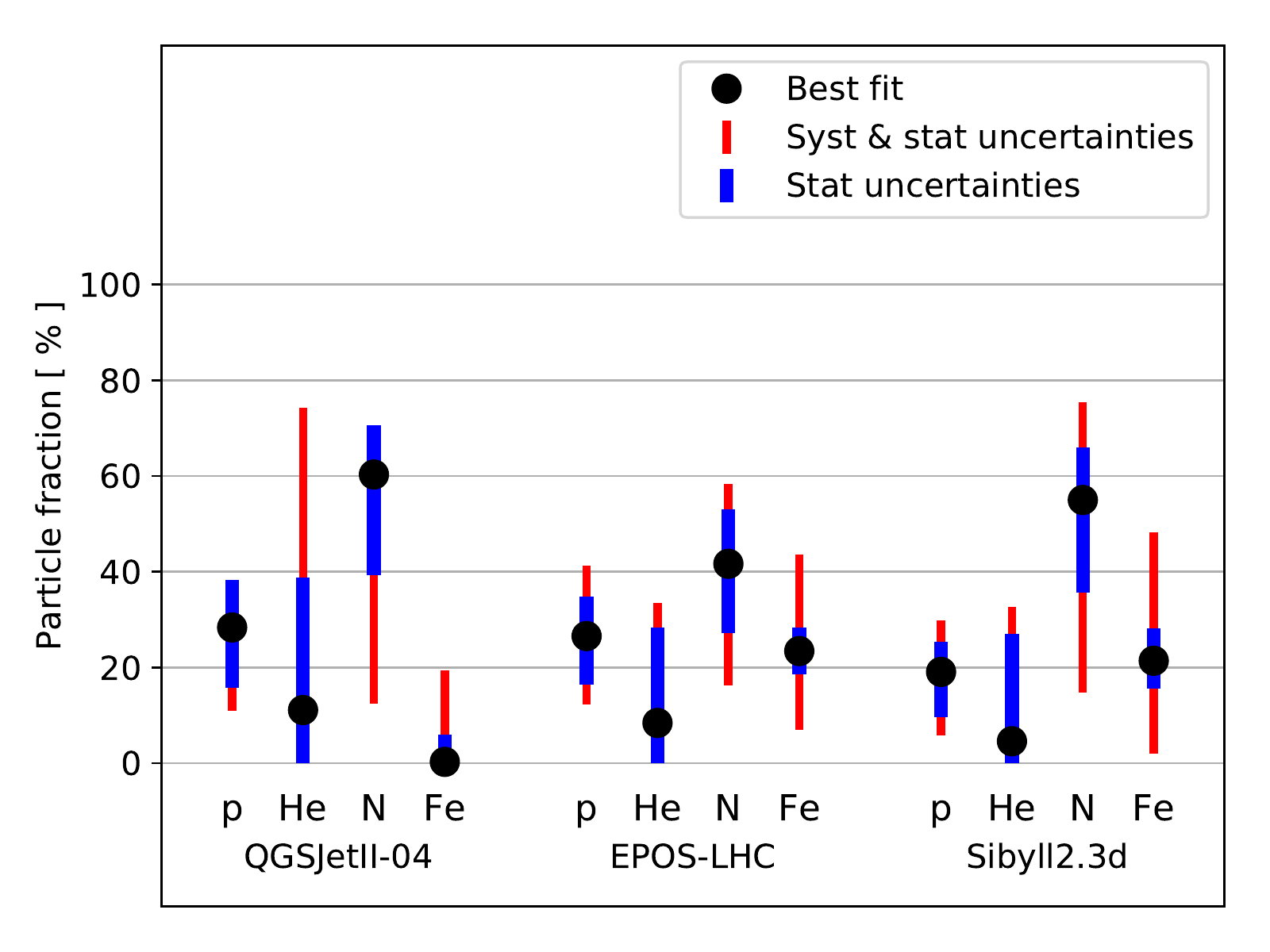}
\caption{Composition results from our dataset, assuming each of the three hadronic interaction models shown at the bottom. The best fit is shown along with statistical and systematic uncertainties.}
\label{fig:composition_results}
\end{center}
\end{figure}

The results for the hadronic interaction models QGSJetII-04, EPOS-LHC, and Sibyll-2.3d are plotted in Fig.~\ref{fig:composition_results}.

For EPOS-LHC, it is seen that the fit favors a more heavy composition, with more iron instead of nitrogen, and with a significant upper bound on the helium fraction. Sibyll-2.3d shows a still somewhat heavier composition, with a lower proton fraction and a higher nitrogen fraction at best fit.
Otherwise, the results are very similar across these three hadronic interaction models, especially considering the uncertainty margins. 
The intermediate-mass elements in the C/N/O range (possibly stretching to somewhat higher mass numbers as well) are dominant, and there is a significant fraction of light elements, i.e.,~protons and helium, at best fit ranging from $23$ to $\unit[39]{\%}$, depending on the interaction model.
Apart from a crossover from C/N/O to iron, the choice of hadronic interaction model has only a limited effect on the best-fit results. 
From the large intermediate-mass contribution it is clear that the composition cannot be described as a two-component mixture of protons and iron.
This is confirmed by a likelihood ratio test with respect to the 4-component model, yielding a p-value $p < 10^{-10}$.
However, a two-component mixture of protons and nitrogen would work for QGSJetII-04 ($p=0.9$).

For this dataset, helium and nitrogen are not fully resolved in the composition model. The fitted values for the helium and nitrogen fraction are highly anti-correlated, which follows from the requirement that all mix fractions sum up to 1.
Helium and nitrogen are a factor 3.5 apart in atomic mass number, whereas the other consecutive elements are a factor 4 apart; constant factors here correspond to a constant shift in $\ln A$, and the mean of the \Xmax-distributions varies by an amount proportional to $\ln A$.
Moreover, helium and nitrogen have two 'neighboring' elements in the composition model, unlike hydrogen and iron. This increases sensitivity to a systematic shift (up or down) in \Xmax.

Importantly, when comparing the current results to the earlier LOFAR results published in \cite{Buitink:2016}, the results are found to be consistent, after various improvements to the analysis setup, the systematic uncertainties, and having a larger dataset.

The previous, smaller dataset allowed for a near-$\unit[100]{\%}$ helium fraction and essentially no protons; this scenario is now very unlikely. 
As shown in Fig.~\ref{fig:contour_p_He}, a similar but somewhat more constrained interchange between protons and helium is still possible. There are astrophysical scenarios where one would see very few protons but a large helium and carbon fraction.
For example, a transition from a Galactic component dominated by Wolf-Rayet supernovae to a particularly strong extragalactic component at these energies could produce a helium fraction near $\unit[60]{\%}$ plus a small proton fraction of about $\unit[5]{\%}$ at our central energy around $2\,\unit[10^{17}]{eV}$ \cite{ThoudamAandA:2016}.
This is still allowed within $\unit[95]{\%}$ confidence limits. 

Comparing the results to those from Pierre Auger Observatory \cite{Petrera:2019}, starting at $\lg (E/\mathrm{eV})=17.25$, agreement is found within statistical and systematic margins. At best fit, their proton fractions are higher, in line with the higher average \Xmax. The difficulty in resolving helium and nitrogen remains, at their higher level of statistics.

Generally, the statistical margins indicate that the analysis would improve with more data. This is no surprise, at a modest number of showers.
However, systematic uncertainties are also important at any level of statistics, as they enlarge the margins of statistical plus systematic uncertainties together.
This is especially evident in the fitted iron fraction, which is well bounded by statistics, but has substantially expanded margins when systematic uncertainties are included. 
Also, looking once more at the substantially overlapping \Xmax-distributions in Fig.~\ref{fig:xmax_gumbel}, it is clear that achieving lower systematic offsets in \Xmax is still important, to improve the resolution of the composition analysis as well as the separability of the element fractions.

\subsection{Analysis in two energy bins}\label{sect:twobins}
\begin{figure}[h]
\begin{center}
\includegraphics[width=0.7\textwidth]{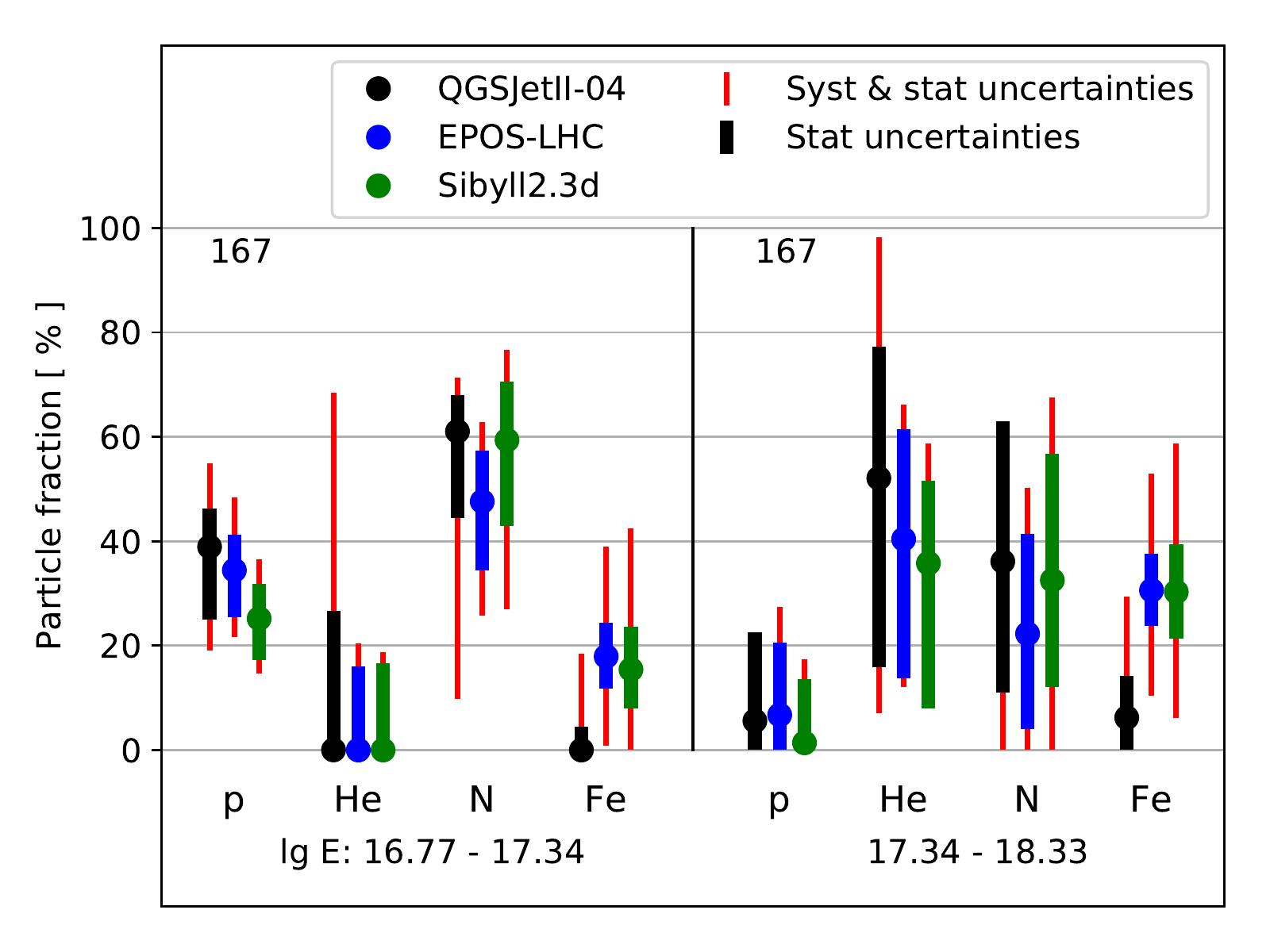}
\caption{Mass composition results in two energy bins. The energy ranges per bin are shown; their main coverage is at $17.14 \pm 0.13$, and $17.65 \pm 0.23$, respectively.}
\label{fig:composition_binned}
\end{center}
\end{figure}

We have divided the dataset into two bins with equal number of showers, being those below versus above the median of $\lg (E/{\mathrm{eV}}) = 17.34$. 
This is a conservative choice with respect to statistical significance.
The results for the three interaction models are shown in Fig.~\ref{fig:composition_binned}.
The coverage of the energy bins, summarized as an average and standard deviation, is mainly at $\lg (E/\mathrm{eV}) = 17.14 \pm 0.13$, and $17.65 \pm 0.23$, respectively.

Notable is the best-fit proton fraction, which is lower in the high-energy bin for all three interaction models.
However, from a likelihood ratio test (see Sect.~\ref{sect:stat_analysis}, bottom) the difference between the two bins was found not to be statistically significant.
That is, a null hypothesis that the observed difference between the two bins would arise from a mass composition constant with energy, as in Fig.~\ref{fig:composition_results}, is not rejected ($p=0.58$).

A similar, simpler test is to use Eq.~\ref{eq:xmax_energy_shift} to take out the first-order energy-dependence of \Xmax, and split the dataset in two equal-sized bins, for lower and higher energy, respectively. This is a non-parametric test which does not depend on element-based composition models using \Xmax-distributions. Also, the elongation rate does not have a strong dependence on the interaction model.

The difference in average $Y_{\rm max}$ between the low and the high-energy bin is not significant ($p=0.25$).
Thus, it is clear that a larger dataset is needed to draw conclusions on possible variations of the element fractions with energy.
Would such a trend towards lower proton fractions be confirmed, it would challenge the hypothesis of a transition from a helium and C/N/O-dominated Galactic to a proton-dominated extragalactic component at energies below $\unit[10^{18}]{eV}$.

\clearpage

\section{Summary}\label{sect:summary}
We have presented an updated cosmic-ray mass composition analysis from LOFAR radio data, built on an improved method for measuring the depth of shower maximum \Xmax of air showers. For the reconstruction of shower parameters, CORSIKA / CoREAS simulations have been used.

We have incorporated several refinements to our analysis, such as including local atmospheric parameters, the Galaxy-based radio calibration and energy measurement, and improved detector characterisation in the fiducial sampling procedure. This leads to an energy resolution of $\unit[9]{\%}$ and an \Xmax-resolution of $\unit[19]{g/cm^2}$ on average, and a systematic uncertainty of $\unit[14]{\%}$ on energy and $\unit[7]{g/cm^2}$ on \Xmax, or $\unit[9]{g/cm^2}$ in the mass composition analysis.

To obtain an unbiased dataset suitable for composition studies, three selection criteria were applied to a set of 720 reconstructed showers, based  on the ensemble of simulations per measured shower. Requiring a shower core precision better than $\unit[7.5]{m}$ gives a sufficient general cut on reconstruction quality. 
We further require that each simulated shower in the CORSIKA / CoREAS ensemble must be able to trigger the LORA particle detector array, and also pass the detection and quality criteria of the LOFAR radio analysis pipeline.
This procedure leaves a sample of 334 showers for analysis of \Xmax-statistics and (mixed) element composition.

The inferred average and standard deviation of \Xmax have been presented in a (log-)energy range of $16.75 < \lg (E/\mathrm{eV}) < 18.25$. The average \Xmax was found to be in line with results from Northern hemisphere-based observatories such as Tunka, Yakutsk, HiRes/Mia, and TALE. However, the values are somewhat lower than those from the Pierre Auger Observatory, where their energy range overlaps. 
The origin of this tension remains unclear and requires additional research.

Apart from the first two moments, we have also analysed the \Xmax-data at (complete) distribution level, using a four-component model of elements, about equally spaced in $\ln A$.
An unbinned maximum likelihood method was found suitable to obtain the best-fitting mass composition in our energy range, together with a separate goodness-of-fit test.
This relies on high-precision parametrizations of the \Xmax-distributions of the elements, as produced by \cite{DeDomenico:2013} and updated by \cite{Petrera:2020}. 

From this analysis, the best-fitting mass composition for our dataset is $\unit[28]{\%}$ protons, $\unit[11]{\%}$ helium, $\unit[60]{\%}$ nitrogen and $\unit[1]{\%}$ iron, assuming the QGSJetII-04 hadronic interaction model. This is averaged over our energy range, with coverage mainly at a log-energy of $\lg (E/\mathrm{eV}) = 17.39 \pm 0.32$.
The EPOS-LHC and Sibyll-2.3d models tend towards a heavier composition, with an iron fraction just above $\unit[20]{\%}$ as a best fit.
The light-mass elements together, protons plus helium, form a fraction of $23$ to $\unit[39]{\%}$ at best fit.
Overall, the differences in composition results between these three important hadronic interaction models are minor, apart from a shift between nitrogen and iron. For these differences to become significant, a larger dataset would be needed.
A division of the dataset into two equal-sized bins, for lower and higher energy respectively, yielded no significant difference, both in the model-based composition analysis and in the difference in elongation-rate-corrected \Xmax (Eq.~\ref{eq:xmax_energy_shift}).

The present results are consistent with the earlier LOFAR results \cite{Buitink:2016}, thus confirming an appreciable light-mass component in this energy range, at a lower level of systematic uncertainties and from an extended dataset. 
The element-based mass composition results are in agreement with those from the Pierre Auger Observatory, within systematic and statistical uncertainties. Hence, the tension in the average \Xmax results does not translate to element-based results outside their uncertainty margins. 
To conclude, we have shown that our \Xmax-analysis per air shower achieves an accuracy in line with the current state of the art, and demonstrates the value of the radio detection method for measuring air showers.

\clearpage

\section*{Acknowledgements}
We acknowledge funding from the European Research Council under the European Union's Horizon 2020 research and innovation programme
(grant agreement n.~640130).
TNGT acknowledges funding from Vietnam National Foundation for Science and Technology Development (NAFOSTED) under grant number 103.01-2019.378. ST acknowledges funding from the Khalifa University Startup grant, project code 8474000237-FSU-2020-13.

LOFAR, the Low Frequency Array designed and constructed by ASTRON, has facilities in several countries, that are owned by various parties (each with their own funding sources), and that are collectively operated by the International LOFAR Telescope (ILT) foundation under a joint scientific policy.

\section*{References}

\bibliographystyle{elsarticle-num}
\bibliography{composition}

\end{document}